\documentclass[12pt,english]{article}
\usepackage[english]{babel}
\usepackage{amsmath,amssymb,amscd}
\usepackage{amsfonts}
\usepackage{graphicx}
\usepackage{appendix}
\usepackage{url}
\usepackage{multirow}
\usepackage{lscape}
\usepackage{array}
\usepackage{booktabs}

\makeatletter
\renewcommand\section{\@startsection {section}{1}{\z@}%
                                   {-5.5ex \@plus -1ex \@minus -.2ex}
                                   {2.3ex \@plus.2ex}%
                                   {\normalfont\large\bfseries}}
\renewcommand\subsection{\@startsection{subsection}{2}{\z@}%
                                     {-3.25ex\@plus -1ex \@minus -.2ex}%
                                     {1.5ex \@plus .2ex}%
                                     {\normalfont\bfseries}}

\numberwithin{equation}{section}
\makeatother
\date{}
\newcommand{\ba}{\begin{align}}

\newcommand{\bea}{\begin{eqnarray}}
\newcommand{\eea}{\end{eqnarray}}
\newcommand{\be}{\begin{equation}}
\newcommand{\ee}{\end{equation}}
\newcommand{\eq}[1]{(\ref{#1})}

\newcommand{\Z}{{\mathbb Z}}

\newcommand{\C}{{\mathbb C}}

\newcommand{\M}{{\mathcal M}}

\def\Tr{{\rm Tr}}

\def\F{\mathbb{F}}

\def\z{\Z}

\newcommand{\cG}{{\mathcal G }}
\newcommand{\cN}{{\mathcal N }}

\newcommand{\cR}{{\mathcal R }}

\newcommand{\ie}{i.e.~}
\def\eq{\begin{equation}}
\def\en{\end{equation}}
\def\eqa{\begin{eqnarray}}
\def\ena{\end{eqnarray}}
\def\ignorethis#1{}

\def\ket#1{| #1\rangle}

\def\VM{V^{\mg}}
\def\Vp{V^\perp}
\def\Vr{V^{\rm rest}}

\def\Vpd{W^\perp}
\def\Vrd{W^{\rm rest}}
\def\vo{V^\perp}

\newcommand{\rt}[1]{\tilde{#1}}

\newcommand{\pl}[1]{\Psi^{#1}}
\newcommand{\pbl}[1]{\bar{\Psi}^{#1}}

\newcommand{\Xl}[1]{\partial X^{#1}}
\newcommand{\Xbl}[1]{\partial\bar{X}^{#1}}

\newcommand{\mg}{\ensuremath{\mathbb{M}_{24}}}

\newcommand{\nc}{\ensuremath{N^{\rm CFT}}}
\newcommand{\vm}{\ensuremath{V^{\mg}}}
\newcommand{\vr}{\ensuremath{V^{\rm rest}}}
\newcommand{\vc}{\ensuremath{V^{\rm CFT}}}

\newcommand{\lo}{\ensuremath{\mathcal{O}}}

\newcommand{\con}[1]{\ensuremath{\boldsymbol{\overline{\textbf{#1}}}}}

\newcommand{\HA}{\ensuremath{H}} 

\newtheorem{thm}{Theorem}

\newcommand{\COne}{1a}
\newcommand{\CTwo}{2a}
\newcommand{\CSeven}{3a}
\newcommand{\CEleven}{2b}
\newcommand{\CTwelve}{4a}

\newcommand{\tg}{T}

\begin{document}

\centerline{\large{Mathieu Moonshine and Symmetry Surfing}}
\bigskip
\bigskip
\centerline{Matthias R.~Gaberdiel$^1$, Christoph A.~Keller$^2$, and Hynek Paul$^1$}
\bigskip
\bigskip
\centerline{$^1$Institute for Theoretical Physics, ETH Zurich}
\centerline{CH-8093 Zurich, Switzerland}
\medskip
\centerline{$^2$Department of Mathematics, ETH Zurich}
\centerline{CH-8092 Zurich, Switzerland}
\bigskip
\bigskip
\begin{abstract}
Mathieu Moonshine, the observation that the Fourier coefficients of the elliptic genus on K3 can be 
interpreted as dimensions of representations of the Mathieu group \mg, has been proven abstractly, but a conceptual
understanding in terms of a representation of the Mathieu group on the BPS states, is missing. 
Some time ago, Taormina and Wendland showed that such an action can be naturally defined
on the lowest non-trivial BPS states, using the idea of `symmetry surfing', i.e., by combining 
the symmetries of different K3 sigma models. In this paper we find non-trivial evidence that
this construction can be generalized to all BPS states.
\end{abstract}
\newpage

\section{Introduction}

Mathieu Moonshine \cite{Eguchi:2010ej} is the observation that the decomposition of the elliptic genus of 
K3 into ${\cal N}=4$ superconformal characters is of the form
\be\label{1.1}
\phi(z,\tau) = 20 \, ch_{\frac{1}{4},0} - 2\, ch_{\frac{1}{4},\frac12} + 90 \, ch_{1,0} + 462\,  ch_{2,0} +  1540\, ch_{3,0} +\cdots \ ,
\ee
where the coefficients appearing in front of the massive representations $ch_{h,0}$ with $h\in\mathbb{N}$ 
correspond to the dimensions of representations $V_h^{\mg}$ of the Mathieu group \mg. 
(Similarly, also the coefficients $20$ and $2$ have an interpretation as virtual representations of \mg.)
This phenomenon is very similar to the case of Monstrous Moonshine~\cite{MR1172696}, 
in that it connects the Fourier coefficients of a modular object --- the elliptic genus of K3
is a weak Jacobi form --- with the representations of a sporadic group.

Initially, this observation was only noted for the first few Fourier coefficients \cite{Eguchi:2010ej}, but by now it is known
to be true for all of them. Indeed, the functional
form of the different twining genera, involving the insertion of an \mg\ group element into the 
elliptic genus, have been found indirectly, and they possess the appropriate modular properties 
\cite{Cheng:2010pq,Gaberdiel:2010ch,Gaberdiel:2010ca,Eguchi:2010fg}.
From this information one can deduce the decomposition of all Fourier coefficients in terms of 
\mg\ representations, and it was shown in \cite{Gannon:2012ck} that all Fourier coefficients  can be 
written in terms of \mg\ representations.

However, while this solves on a certain level the problem of proving Mathieu Moonshine, the 
argument is rather indirect and it does not explain the emergence of \mg\ in this context. 
Indeed, by analogy with Monstrous Moonshine, one would 
expect that there should be something like a vertex operator algebra or conformal field
theory with automorphism group \mg\  whose elliptic genus reproduces the elliptic genus of K3. 
\smallskip

One natural candidate for such a conformal field theory is obviously a sigma model on K3 itself. Unfortunately,
as was shown in \cite{Gaberdiel:2011fg}, no single sigma model contains \mg\ as an automorphism group. 
More specifically, it was shown there that a group $G$ can only be the symmetry group of a K3 sigma model if
it is a subgroup of $Co_1 < Aut(\Lambda)$ whose fixed point
lattice $\Lambda^G$ has at least rank 4. All such groups
and their fixed point lattices are listed in table~1 
of \cite{MR3438323}.
The largest such group is $\Z_2^8: M_{20}$ and has
order 245760, and  the corresponding sigma
model was constructed in \cite{Gaberdiel:2013psa}.
In any case, it follows from the arguments of \cite{Gaberdiel:2011fg} that an VOA or conformal field theory 
with \mg\ symmetry cannot just come form a single K3 sigma model. 

In \cite{Taormina:2013mda,Taormina:2013jza} it was proposed that one may attempt to
circumvent this no-go-theorem by combining the symmetries from different points of the moduli space
of K3 sigma models. In particular, they considered three different $\mathbb{T}^4/\mathbb{Z}_2$
orbifold points, and showed that the \mg\ action on $V_1^{\mg}$, that is for 
$\mathbf{90}= \mathbf{45}\oplus \mathbf{\overline{45}}$, could be reconstructed from these three symmetry groups. 
In this paper we attempt (and partially succeed) in generalizing their construction to all $h$. 
Even though we do not actually construct the full representations, we find very non-trivial evidence 
indicating that this procedure will actually work. 
\smallskip

The basic idea behind this approach is that the elliptic genus only counts 
supersymmetric states, more precisely states that are
at least $\frac{1}{4}$ BPS. Since the elliptic genus does not know about the non-BPS states,
it is not necessary that $\mg$ is a symmetry of them; it is enough
if $\mg$ is a symmetry of the subspace of BPS states
\be
V^{\rm BPS} \subset \mathcal{H}\ .
\ee
At a generic point of moduli space $\M$ of K3, we expect the coefficients of the elliptic genus 
to agree precisely with the graded dimensions $V_h^{\rm BPS}$ (possibly up to some signs and multiplicities
coming from the right-moving Ramond ground states). However, at such a generic point, the 
automorphism symmetry of the corresponding K3 sigma model is typically trivial. 

We shall therefore consider the torus orbifolds $\mathbb{T}^4/\mathbb{Z}_2$, that describe a 
$16$-dimensional sub-locus of the full $80$-dimensional moduli space $\M$. For every sigma model on this moduli space
the graded dimensions $\dim V_h^{\rm BPS}$ (for $h>1$) are always bigger than the corresponding coefficients
in the elliptic genus --- the elliptic genus is a partial index (effectively the Witten index
with respect to the right-movers), and hence contributions can (and will in fact) cancel. 
What is special about the $\mathbb{Z}_2$ torus orbifolds is that their BPS spectrum is essentially independent
of which orbifold one considers: there are BPS states from the zero-momentum zero-winding
part of the untwisted sector, as well as from the $16$ twisted sectors. (For special sizes of the torus
there may be additional BPS states with non-zero momentum and winding, but we shall avoid these 
special points.) Furthermore, one 
can argue that the cancellation that appears in the calculation of the elliptic genus, will only
involve the untwisted sector states, as well as {\em one} of the $16$ twisted sectors. In particular,
we therefore know on general grounds that the BPS states that arise from $15$ of the $16$ twisted
sectors will form a subspace $\Vp_h \subset V_h^{\mg}$. Furthermore, there are three special $\mathbb{Z}_2$
orbifolds corresponding to the square Kummer surface $X_0$ with symmetry group $G_0:=(\z_2)^4\rtimes(\z_2\times\z_2)$, the 
tetrahedral Kummer surface $X_1$ with symmetry group $G_1:=(\z_2)^4\rtimes A_4$ , and the triangular 
Kummer surface $X_2$ with symmetry group $G_2:=(\z_2)^4\rtimes S_3$~\cite{Taormina:2013mda,Taormina:2013jza}. 
Thus we can read off the action of  the three geometric symmetry groups $G_i$ on $\Vp$, and hence
construct their representations 
$\rho^{i} : G_i \rightarrow Aut(\Vp)$ explicitly.

One of the central results of \cite{Taormina:2013mda,Taormina:2013jza}
is that the geometric groups can be naturally combined into the so-called octad group $G=(\z_2)^4\rtimes A_8$, 
which is a maximal subgroup of \mg.
One should then expect that the octad group $G$ acts naturally on each $\Vp_h$, and this will indeed turn out 
to be the case. Indeed, starting from the known decomposition of $\VM_h$ in terms of \mg\ representations, we
can consider the branching under the octad group $G\subset \mg$, and regard $\VM_h$ as a representation of $G$. 
Then there is a simple group theoretic rule for which representations of $G$ arise in $\Vp$ (and which account for the 
rest of the spectrum $\Vr$ --- in fact the representations that appear in $\Vr$ are precisely those that are invariant
under the normal factor group $\tg:=(\z_2)^4$.) This prediction can then be tested by explicit comparison 
with the geometric actions $\rho^{i}$ on $\Vp_h$, and we find beautiful agreement. 
More explicitly we prove the following theorem:
\begin{thm}\label{thm1}
We can decompose the $\mg$ representations $\VM_h$
into representations $\Vp_h$ and $\Vr_h$ of the octad group $G=(\z_2)^4\rtimes A_8 \subset \mg$,
$$\VM_h =  \Vp_h\oplus \Vr_h $$
where $\Vr_h:=V_h^{(\z_2)^4}$ is the subspace
invariant under the normal factor group $\tg:=(\z_2)^4$. 
$\Vp$ then has the properties:
\begin{enumerate}
\item The representation $\rho_h : G \rightarrow Aut(\Vp_h)$  
is induced from a representation $\pi_h$ of the subgroup $\HA < G$, 
where 
$\HA=((\z_2)^6\rtimes PSL(3,2))\rtimes\z_2$ is the stabilizer subgroup of one of the $16$ twisted sectors. 
\item
The  restriction of $\rho_h$ to any of the three geometric subgroups
$G_i$, $i=0,1,2$,  is isomorphic to the representations $\rho^{i}_h$,
\be\label{Thm1}
\mathrm{Res}_{G_i}\rho_h \cong \rho_h^{i} \ .
\ee 
\end{enumerate}
\end{thm}
Note that for the $h=1$ case considered in \cite{Taormina:2013mda,Taormina:2013jza},
$\Vr_1 = 0$; however, for $h>1$, $\Vr_h\neq 0$. We should mention that the contribution from $\Vp$ accounts
precisely for the contribution from $15$ of the $16$ twisted sectors; in particular, the dimensions
of the octad representations add up correctly to the corresponding coefficients.

\smallskip

The paper is organized  as follows. In section~\ref{s:VM24} we introduce our notations, and construct
the spaces $\VM$ and $\Vp$. In section~\ref{s:Gact} we discuss the 
action of $G$ and the geometric subgroups $G_i$. In particular
we define the stabilizer subgroup $H$ and the geometric representations
$\rho^{i}$, and we explain why Theorem~\ref{thm1} gives evidence
for the symmetry surfing proposal.
In section~\ref{s:Results} we finally prove Theorem~\ref{thm1}, and section~\ref{s:conclusions} contains
our conclusions.

\section{The elliptic genus of $\mathbb{T}^4/\Z_2$}\label{s:VM24}
\subsection{The elliptic genus of K3}
\label{sec:elliptic_genus}

The elliptic genus of K3 is defined as the trace over the ${\rm R}\rt{{\rm R}}$ sector of the sigma model
\begin{equation}
\label{ellgen}
	\phi_{\rm K3}(\tau,z) = \text{Tr}_{{\rm R}\rt{{\rm R}}}\left(q^{L_0-\frac{1}{4}}y^{J_0}(-1)^F\bar{q}^{\rt{L}_0-\frac{1}{4}}(-1)^{\rt{F}}\right)\ ,
\end{equation}
where we denote the right-movers by a tilde. 
As usual we take $q:=\text{exp}(2\pi i\tau)$ and $y:=\text{exp}(2\pi iz)$, where $\tau\in\mathbb{H}$ and $z\in\mathbb{C}$;
furthermore, $F$ and $\rt{F}$ are the fermion number operators.
Although also the anti-holomorphic variable $\bar{q}=\text{exp}(-2\pi i\bar{\tau})$ enters its definition, the elliptic genus itself is 
actually independent of $\bar{\tau}$, since it effectively computes the Witten index for the right-moving states.
It is therefore a topological quantity and independent of the specific K3 surface chosen.
In~\cite{Eguchi:1988vra}, the elliptic genus of K3 was explicitly calculated to be
\begin{align}
\begin{split}
\label{ellgentheta}
	\phi_{\rm K3}(\tau,z) &= 8 \left[\left(\frac{\vartheta_2(\tau,z)}{\vartheta_2(\tau,0)}\right)^2 + \left(\frac{\vartheta_3(\tau,z)}{\vartheta_3(\tau,0)}\right)^2 + \left(\frac{\vartheta_4(\tau,z)}{\vartheta_4(\tau,0)}\right)^2\right]\\
										&=2 y+20+\frac{2}{y} + q \left(20 y^2-128 y +216 -\frac{128}{y}+\frac{20}{y^2}\right) + \lo(q^2)\ ,
\end{split}
\end{align}
where $\vartheta_i(\tau,z)$ are the usual Jacobi theta functions.
Because of its representation as a sum of Jacobi theta functions, the elliptic genus has nice transformation properties under the modular group. 
In particular, $\phi_{\rm K3}(\tau,z)$ can be identified with a weak Jacobi form of weight zero and index one, which also follows from general 
string theory arguments~\cite{Kawai:1993jk}. Using the modular properties of Jacobi theta functions 
one finds
\begin{equation}
\begin{alignedat}{2}
\label{invariance}
	&S: \qquad\phi_{\rm K3}(-\frac{1}{\tau},\frac{z}{\tau}) &&=e^{\frac{2\pi i z^2}{\tau}}\phi_{\rm K3}(\tau,z)\ ,\\
	&T: \qquad\phi_{\rm K3}(\tau+1,z) &&=\phi_{\rm K3}(\tau,z)\ .
\end{alignedat}
\end{equation}

In what follows, it will be more natural to work in the 
NS$\rt{{\rm R}}$ sector. Because of the underlying ${\cal N}=2$ superconformal symmetry, we can go back and forth
between the NS$\rt{{\rm R}}$ and the R$\rt{{\rm R}}$ sector by spectrally flowing  the left-movers, i.e., by 
$y \mapsto \sqrt{q} y$. We also want to remove the left moving $(-1)^F$ insertion, which we can do by replacing
$y\mapsto -y$. The elliptic genus in the NS$\tilde{\rm R}$ sector is thus defined as
\be
\label{nsgen}
\phi_{\rm K3}^{\rm NS}(\tau,z)= \text{Tr}_{{\rm NS\rt{R}}}\left(q^{L_0-\frac{1}{4}}y^{J_0}\bar{q}^{\rt{L}_0-\frac{1}{4}}(-1)^{\rt{F}}\right)\ ,
\ee
which for K3 gives
\begin{align}
\begin{split}
\label{nsgenus}
\phi_{\rm K3}^{\rm NS}(\tau,z)  &= 8\left[-\left(\frac{\vartheta_4(\tau,z)}{\vartheta_2(\tau,0)}\right)^2-\left(\frac{\vartheta_1(\tau,z)}{\vartheta_3(\tau,0)}\right)^2+\left(\frac{\vartheta_2(\tau,z)}{\vartheta_4(\tau,0)}\right)^2\right] \\
				&= q^{-\frac{1}{4}}\left[-2 + \sqrt{q}\left( 20y+{20\over y}\right) + q \left( -2y^2+128-{2\over y^2}\right) + \lo(q^{3\over2})\right]\ .
\end{split}
\end{align}
Note that only states whose right-moving part are Ramond ground states contribute
to the elliptic genus. We can label those right-moving ground states by their ${\rm SU}(2)_R$ spins $l=0$ and $l=\frac12$.
(Here  ${\rm SU}(2)_R$ is the $R$-symmetry, i.e., the ${\rm SU}(2)$ subalgebra of the ${\cal N}=4$ superconformal algebra.)
Their Witten index is $1$ and $-2$, respectively.

\subsection{The elliptic genus of $\mathbb{T}^4/Z_2$}
\label{sec:partition}

As we have already mentioned before, the elliptic genus is a partial index which is independent of the specific point in moduli 
space of K3 sigma models. We may therefore choose the K3 surface limit of a $\mathbb{T}^4/\z_2$ orbifold as our target space.
In this section we want to describe the full space of states $W^{\rm CFT}$ of the corresponding conformal field theory; we shall
also denote by $\vc$ the subspace of primary states. Both $W^{\rm CFT}$ and $\vc$ are graded by conformal weight $h$.

Let us begin by introducing some notation. We denote the complex fermions and complex bosons of $\mathbb{T}^4$
as 
\be
\pl{i} \ , \quad \pbl{i}  \ , \qquad  \Xl{i}\ , \quad \Xbl{i}  \ ,
\ee
where $i=1,2$; we also sometimes denote these fields collectively by $\Psi^I$ and $\partial X^I$ with
$I=1,2,3,4$. For the right-movers we use the same convention, except that they are decorated by a tilde. 
The $\cN=4$ superconformal algebra can then be constructed in the usual way.
The symmetry at the torus orbifold group is actually bigger. In particular 
there is an additional ${\rm SU}(2)$ symmetry which commutes with the $\cN=4$ . 
Under this symmetry the fields transforms as doublets,
\be\label{su2}
{\rm SU}(2) : \quad \begin{array}{c} \Psi\ , \  \partial X\qquad \bf{2}\\ \bar\Psi\ , \  \partial \bar X \qquad \bar{\bf{2}} 
\end{array}
\ee
The details of this are described in appendix~\ref{app:N4}.
This symmetry is crucial in what follows since we can think of the \mg\ 
action as being associated to group elements in this ${\rm SU}(2)$.

As usual, the $\z_2$ orbifold acts on the fields as
\begin{equation}
\label{orbifold}
	\Psi^I(z) \longmapsto -\Psi^I(z)\ , \qquad \partial X^I(z) \longmapsto -\partial X^I(z)\ ,
\end{equation}
and similarly on the right-movers. In order to specify its action on the Hilbert space, 
it remains to understand the behavior of the ground states. For this it is convenient to work in the 
R$\rt{\rm R}$ sector, and flow to NS$\rt{\rm R}$ at the end.

\label{sec:groundstates}
Let us start with the untwisted sector.
Here both the bosonic $\partial X^I(z)$ and fermionic fields $\Psi^I(z)$ have integer-valued modes. 
We thus have four fermionic zero modes $\Psi_0^I$ , and similarly four right-moving zero modes $\widetilde{\Psi}_0^I$
such that the theory has degenerate charged ground states. 
Let us define the $\pbl{i}_0$ to be the annihilation operators,
and the $\pl{i}_0$ to be the creation operators, so that the highest weight state
$\ket{0}_{\rm R}$ satisfies
\begin{equation}
\label{annih}
	\bar{\Psi}_0^i\ket{0}_{\rm R}=0\ .
\end{equation}
The four left-moving ground states thus read
\begin{equation}
\begin{alignedat}{2}
	&\ket{0}_{\rm R}\ ,	\qquad&&\Psi_0^1\Psi_0^2\ket{0}_{\rm R}\ , \label{2.10} \\
	&\Psi_0^1\ket{0}_{\rm R}\ ,\qquad &&\Psi_0^2\ket{0}_{\rm R}\ .
\end{alignedat}
\end{equation}
They are charged under the SU$(2)_R$ current with charges $l=0,\frac{1}{2}$, as one can check by applying $J_0^3$. Therefore, the four ground states decompose into the ground states of $l=0$ and $l=\frac{1}{2}$ representations as
\begin{equation}
\begin{alignedat}{2}
\label{rvacua}
	2\cdot&\ket{l=0}_{\rm R}:\qquad&&~\Psi_0^1\ket{0}_{\rm R}\ ,~\Psi_0^2\ket{0}_{\rm R}\ ,\\
	&\ket{l=\tfrac{1}{2}}_{\rm R}:&&\ket{0}_{\rm R}\ ,~\Psi_0^1\Psi_0^2\ket{0}_{\rm R}\ .
\end{alignedat}
\end{equation}
The situation for the right-moving part is completely analogous.

Now we need to take the $\z_2$ orbifold projection into account. We choose the highest weight state $\ket{0}_{\rm R}\otimes  \ket{0}_{\rt{\rm R}}$ 
to be orbifold even. Remember that the $\cN=4$ superconformal generators, as stated in equations~\eqref{Jdef} -- \eqref{Gdef2}, are bilinear in the 
fields and therefore are $\z_2$ invariant. For the different untwisted ground states from equation~\eqref{rvacua} this results in 
$\ket{l=0}_{\rm R}$ being orbifold odd and $\ket{l=\tfrac{1}{2}}_{\rm R}$ even.

It will be convenient to spectrally flow the left-moving states to the NS sector. 
Since spectral flow maps $l=0$ to $l=\frac{1}{2}$ and vice versa, we have
\begin{equation}
	\ket{l=\tfrac{1}{2}}_{\rm R} \longmapsto \ket{l=0}_{\rm NS} \ ,
\end{equation}
and thus $\ket{0}_{\rm NS}\otimes\ket{l=\tfrac{1}{2}}_{\rt{\rm R}}$ is orbifold even. All in all, the untwisted ground states before the 
$\z_2$ orbifold projection decompose therefore into
\begin{equation}
\label{untwvacuum}
	\ket{0}_{\rm NS}\otimes\ket{l=\tfrac{1}{2}}_{\rt{\rm R}} ~\oplus~ 2\cdot(\ket{0}_{\rm NS}\otimes\ket{l=0}_{\rt{\rm R}})\ ,
\end{equation}
where the first term is orbifold even, while the second one is orbifold odd. The factor of two accounts 
for the two right-moving $l=0$ ground states $\widetilde{\Psi}^i_0\ket{0}_{\rt{\rm R}}$, $i=1,2$. 
Let us denote by $U_{l=\frac12}(q,y)$ the partition function of all orbifold even left-moving descendants of 
$\ket{0}_{\rm NS}\otimes\ket{l=\frac{1}{2}}_{\rt{\rm R}}$, and by $U_{l=0}(q,y)$ the partition function 
of all orbifold even (left-moving) descendants of $\ket{0}_{\rm NS}\otimes\ket{l=0}_{\rt{\rm R}}$. Explicit expressions
for $U_{l=0}$ and $U_{l=\frac12}$ are given in appendix~\ref{app:ellgenus}.

In the twisted sector, the states are localized at fixed points of the $\mathbb{T}^4/\z_2$ orbifold. There are $2^4$ fixed points and thus we have 16 corresponding ground states which we denote by $\ket{\alpha}$, $\alpha\in\F_2^4$.
In the twisted R$\rt{\rm R}$ sector, there are no fermionic zero modes, so that there is
indeed just a single ground state, and we have
$l=0$. We choose those ground states to be orbifold even. 

Once we flow to the left-moving NS sector, the fermions become integer moded, and
we introduce fermionic zero modes. By the analogous construction as above, 
the twisted ground states are given by
\begin{equation}
\label{twvacuum0}
	\ket{l=\tfrac{1}{2}}\hspace{-5pt}\left.\right._{\rm NS}^\alpha\otimes\ket{l=0}_{\rt{\rm R}}^\alpha ~
	\oplus~ 2\cdot(\ket{l=0}_{\rm NS}^\alpha\otimes\ket{l=0}_{\rt{\rm R}}^\alpha)\ ,
\end{equation}
where the label $\alpha$ denotes one of the 16 fixed points. 
We will denote by $T_{l=0}^\alpha(\tau,z)$ the partition function of all 
orbifold even descendants. An explicit expression can again be found in
appendix~\ref{app:ellgenus}.

In total, we can thus write the NS elliptic genus of the orbifold theory as
\begin{equation}
\label{together}
	\phi_{\rm K3}^{\rm NS}(\tau,z) = -2~U_{l=\frac{1}{2}}(\tau,z) + U_{l=0}(\tau,z) + \sum_{\alpha=1}^{16} T_{l=0}^\alpha(\tau,z)\ .
\end{equation}
Here $U_{l=\frac12}$ is the contribution of the untwisted states
whose right-moving Ramond ground state is in the $\cN=4$ representation with $l=1/2$,
which explains the factor of $-2$ coming from the specialization $\bar{z}=0$.
$U_{l=0}$ is the contribution of the untwisted states whose right-moving
Ramond states are in the $l=0$ representation, and $T_{l=0}$ is the contribution
of a single twisted sector.
In particular they can all be decomposed
into $\cN=4$ representations as
\begin{align}
\label{udecompB}
		U_{l=\frac{1}{2}}(\tau,z) &= ch_{0,l=0}^{\rm NS}(\tau,z) + \sum_{n=1}^\infty{ B_n~ch_{h=n,l=0}^{\rm NS}(\tau,z)}\ ,\\
		\label{udecompC}
		U_{l=0}(\tau,z) &= 	4~ch_{0,l=\frac{1}{2}}^{\rm NS}(\tau,z) + \sum_{n=1}^\infty{ C_n~ch_{h=n,l=0}^{\rm NS}(\tau,z)}\ ,\\
		\label{udecompD}
		T_{l=0}^\alpha(\tau,z) &= 	ch_{0,l=\frac{1}{2}}^{\rm NS}(\tau,z) + \sum_{n=1}^\infty{ D_n~ch_{h=n,l=0}^{\rm NS}(\tau,z)}\ ,
\end{align}
with
\begin{align}
\begin{split}
\label{ucoeff0}
		B_n &= \{3,~1,~18,~15,~68,~89,~249,~358,~799,~1236,\ldots \}\ ,\\
		C_n &= \{0,~16,~8,~72,~80,~264,~360,~904,~1360,~2808,\ldots \}\ , \\
		D_n &= \{6,~28,~98,~282,~728,~1734,~3864,~8182,~16618,\ldots\}\ . 
\end{split}
\end{align}
Accordingly, the number of contributing primary states $A_n$, and thus the dimensions of $\vc_h$, can be computed at every level by
\begin{equation}
\label{contributingprim}
	\dim\left(\vc_n\right) = A_n = -2~B_n + C_n + 16~D_n\ .
\end{equation}
Mathieu Moonshine then means that the $A_n$ can be written as 
(simple) combinations of dimensions of irreducible representations 
of $\mg$.


\section{Symmetry surfing for torus orbifolds}\label{s:Gact}
\subsection{The geometric groups}\label{geomgroup}

In the analysis of Taormina and Wendland the 
$24$-dimensional even self-dual  Niemeier lattice $N$ that is uniquely characterised 
by the property to contain the root lattice $A_1^{24}$ plays a preferred role. The reason
for this is that the symmetry groups of Kummer surfaces, i.e., orbifolds $\mathbb{T}^4/\Z_2$, 
can all be described as subgroups of the automorphism group of $N$. All relevant 
automorphisms are permutations in $S_{24}$ (permuting the
$24$ $A_1$ factors of the root sublattice). Since they must furthermore 
preserve the Golay code $\cG_{24}$ (that describes the additional `glue' vectors 
that are present in the self-dual lattice $N$), they are automatically elements 
of \mg. (Indeed, \mg\ can be defined as the subgroup of $S_{24}$ that leaves
the Golay code invariant.)

Any Kummer surface will always have at least the translational
subgroup $\tg:=\Z_2^4$ as a symmetry group. As elements of $S_{24}$,
its generators are given by
\begin{align}
\begin{split}
\iota_1&=(1,11)(2,22)(4,20)(7,12)(8,17)(10,18)(13,21)(14,16)\ ,\\
\iota_2&=(1,13)(2,12)(4,14)(7,22)(8,10)(11,21)(16,20)(17,18)\ ,\\
\iota_3&=(1,14)(2,17)(4,13)(7,10)(8,22)(11,16)(12,18)(20,21)\ ,\\
\iota_4&=(1,17)(2,14)(4,12)(7,20)(8,11)(10,21)(13,18)(16,22)\ .
\end{split}
\end{align}
Depending on the shape of the torus, the symmetry group can be bigger
than that. There are three particular Kummer surfaces with larger symmetry
groups that we are interested in.
These are the square Kummer surface $X_0$ with symmetry group $G_0:=(\z_2)^4\rtimes(\z_2\times\z_2)$, 
the tetrahedral Kummer surface $X_1$ with symmetry group $G_1:=(\z_2)^4\rtimes A_4$, and the 
triangular Kummer surface $X_2$ with symmetry group $G_2:=(\z_2)^4\rtimes S_3$~\cite{Taormina:2013mda,Taormina:2013jza}. 
We denote the generators for the non-translational part of the symmetry groups, following 
\cite{Taormina:2013mda,Taormina:2013jza} as follows: $\alpha_1,\alpha_2$ for $G_0$; 
$\gamma_1,\gamma_2,\gamma_3$ for $G_1$; and $\beta_1,\beta_2$ for $G_2$. In terms of permutations they are given by
\begin{align}
\begin{split}
\alpha_1&=(4, 8)(6, 19)(10, 20)(11, 13)(12, 22)(14, 17)(16, 18)(23, 24)\ ,\\
\alpha_2&=(2, 21)(3, 9)(4, 8)(10, 12)(11, 14)(13, 17)(20, 22)(23, 24)\ ,\\
\beta_1&=(2, 17, 14)(4, 7, 8)(10, 16, 12)(11, 13, 21)(18, 20, 22)(5, 24, 23)\ ,\\
\beta_2&=(2, 21)(3, 9)(4, 8)(10, 12)(11, 14)(13, 17)(20, 22)(23, 24)\ ,\\
\gamma_1&=(2, 8)(7, 18)(9, 24)(10, 22)(11, 13)(12, 17)(14, 20)(15, 19)\ ,\\
\gamma_2&=(2, 18)(7, 8)(9, 19)(10, 17)(11, 14)(12, 22)(13, 20)(15, 24)\ ,\\
\gamma_3&=(2, 12, 13)(4, 16, 21)(7, 17, 20)(8, 22, 14)(9, 19, 24)(10, 11, 18)\ .
\end{split}
\end{align}

This presentation of the symmetry groups $G_i$ is particularly well-adapted
for the symmetry surfing philosophy. In particular it is straightforward to
combine them into an overarching symmetry group $G$ by combining
all the generators.
The resulting group is the so-called octad group $G=(\z_2)^4\rtimes A_8$. It can
be described as a maximal subgroup of $\mg$ obtained by the setwise stabilizer of a particular 
``reference octad'' in the Golay code, which we take to be  $O_9 = \{3,5,6,9,15,19,23,24\}\in \cG_{24}$. 
The octad subgroup is of 
order $322560$, and its index in $\mg$ is 759, which is precisely the number of different reference 
octads one can choose.
Note in particular that it is too big to fit in the no-go
theorem of \cite{Gaberdiel:2011fg}, and that we could not have gotten this group from
the symmetries of a single K3 sigma model.

\subsection{Marginal deformations and $\Vp$ and $\Vr$ }
For theorem~\ref{thm1} we decomposed $\VM$ into $\Vp$ and $\Vr$. Let us here motivate this decomposition from a physical point of view, and argue that it is natural from the point of view of symmetry surfing.

It is obvious from the discussion of the orbifold spectrum that the BPS spectrum
(for generic tori) comes from the zero-momentum zero-winding part of the untwisted
sector, as well as from the $16$ twisted sectors. However, since all $B_n>0$, see eq.~(\ref{ucoeff0}),
it follows from eq.~(\ref{contributingprim}) that there are always (i.e., for generic
$\mathbb{Z}_2$ torus orbifolds) non-trivial cancellations in the elliptic genus. The space of BPS states $V^{BPS}$ of a tours orbifold is thus too big for $\VM$, and we need a way to pick out a subspace of $V^{BPS}$, which we can then try to identify with $\VM$. 

At a generic point in the moduli space, we should expect there to be no non-trivial
cancellations, i.e., the Fourier coefficients of the elliptic genus should agree with the dimensions
of the corresponding space of BPS states.
We can try to move away from the special Kummer surfaces to a more generic point in the moduli space
by perturbing the conformal field theory by an exactly marginal operator $\Phi$. This will lift a subset of the BPS states in $V^{BPS}$. In fact, if $\Phi$ is generic enough, we would expect that all BPS states that can be lifted, will be lifted. We can then try to define $\VM$ as the subspace of states in $V^{BPS}$ at the original torus orbifold point that are not lifted under the perturbation $\Phi$. 

As we will explain in section~\ref{ss:geoact}, for any torus orbifold the translation group $T=\Z_2^4$ is a symmetry of the theory that permutes the 16 twisted sectors. More precisely, the $16$-dimensional space of twisted sector ground states decomposes into a trivial representation and a 15-dimensional representation --- the $15$-dimensional representation is obviously not irreducible since all irreducible representations of $T=\Z_2^4$ are one-dimensional, but it only contains non-trivial representations of $T=\Z_2^4$. (In fact, it contains each such representation exactly once.) In the notation of section~\ref{ss:geoact}, the invariant linear combination of twisted sectors is given by $N_{0}^{\rm CFT}$, and we choose $\Phi$ to be a marginal operator coming from $N_{0}^{\rm CFT}$.
This choice implies that the BPS states coming from the $15$-dimensional
space of twisted sectors orthogonal to $\Phi$ --- these are the states that transform in the $15$-dimensional representation of $T=\Z_2^4$ mentioned above --- will be unaffected. To see this, we note that  that twisted sectors only have right-moving $l=0$ BPS states, see eq.~(\ref{twvacuum0}). To lift them, we thus need to combine them with right-moving $l=1/2$ BPS states, which necessarily come from the untwisted sector, see eq.~(\ref{untwvacuum}). 
On the other hand, because of the representation theory of the translation group $T=\Z_2^4$, the perturbation by the singlet state from $N_{0}^{\rm CFT}$ cannot mix a BPS state from the untwisted sector (that also transforms trivially under the translation group) with a twisted sector state transforming in the $15$-dimensional representation of $T=\Z_2^4$ from above. Indeed, the correlation functions, where $\phi_{\rm ut}$ is from the zero-momentum and zero-winding  part of the untwisted sector and $\phi_{\rm tw}$ is a twisted sector state in the $15$-dimensional representation of $T=\Z_2^4$ 
\be\label{pertlift}
\langle \phi_{\rm ut}\phi_{\rm tw}\underbrace{\Phi\ldots \Phi}_n \rangle  = 0 
\ee
vanish for all $n$, and hence to arbitrary order in perturbation theory, no such mixing can occur.
Thus we expect that the
$15$-dimensional space orthogonal to $\Phi$ (that defines the $15$-dimensional representation of $T=\Z_2^4$) will form a subspace of surviving BPS states
which we will call $\Vp$. In the untwisted sector and the 
remaining twisted sector, we expect all non-generic BPS
states to be lifted, so that there are no more cancellations
in the elliptic genus; we will call the space of BPS states that survive $\Vr$.
In particular this requires that the linear combination
of coefficients 
\be
\dim( \Vr_n) \equiv - 2 B_n + C_n + D_n \geq 0 \ . 
\ee
This physical argument thus motivates us to define
\be
V^\mg = \Vr \oplus \Vp\ ,
\ee
where $\Vr$ is indeed the subspace invariant under $T=\Z_2^4$, and $\Vp$ is the orthogonal complement.
Note that we know $\Vp$ explicitly, as it is simply
the direct sum of 15 twisted sectors. To describe
$\Vr$ explicitly, on the other hand, we would have 
to compute which states are lifted. We will not do this here,
and instead concentrate on $\Vp$.

\subsection{Action of the geometric groups $G_i$ on $\Vp$}\label{ss:geoact}
Let us now discuss the action of the $G_i$ on $\Vp$.
We first discuss how the geometric groups $G_i$ act at their respective points in moduli space and construct their representations $\rho^i$.
For concreteness, we fix the ground state $|\alpha\rangle$ such that
\be
\pbl{i}_0|\alpha\rangle = 0 \ .
\ee
From the discussion around eqs.~(\ref{2.10}) and (\ref{twvacuum0}), 
we know that there are two ground states 
$|\alpha\rangle$ and $\pl{1}_0\pl{2}_0|\alpha\rangle$
which survive the orbifold projection, which in fact
form the $l=1/2$ doublet of the $R$-symmetry ${\rm SU}(2)_R$. 
Since the group action is supposed to commute with the ${\cal N}=4$ superconformal algebra, both these states transform the same
way under $G$. We can thus concentrate on the state $|\alpha\rangle$ only.
Following the construction in~\cite{Taormina:2013mda}, we introduce linear combinations of these twisted ground states, given by
\begin{equation}
\label{ndef}
	N_{\alpha}^{\rm CFT}:=\frac{1}{4} \sum_{\beta \in  \mathbb{F}_2^4}(-1)^{\left\langle \alpha,\beta \right\rangle}~\ket{\beta}\ ,
\end{equation}
where $\left\langle \cdot,\cdot \right\rangle$ is the standard scalar product on $\F_2^4$. The translational subgroup $\tg = \mathbb{Z}_2^4$ 
of the $G_i$ acts on $N_{\alpha}^{\rm CFT}$ via the map
\begin{equation}
\label{z24rep}
	\iota_{\beta}(N_{\alpha}^{\rm CFT}) = (-1)^{\left\langle \beta,\alpha \right\rangle}~N_{\alpha}^{CFT}\ , \qquad \forall~\beta\in\F_2^4\ .
\end{equation}
It is clear from the definition \eqref{ndef} that the perturbing field $\Phi$ sits in the $N_{0000}^{\rm CFT}$ sector; 
the $15$-dimensional space of twisted sector (ground) states that are unaffected by the perturbation is then spanned
by the remaining 15 $N_{\alpha}^{\rm CFT}$ ($\alpha\neq0$). These ground states carry a faithful representation of $G$,
which we identify with the $15$-dimensional representation $\mathcal{A}$ of Aff$(\mathbb{F}_2^4)$. In~\cite{Taormina:2013mda}, 
an isomorphism of $\tg$-representations between $\mathcal{A}$ and Margolin's base $\mathcal{B}=\{N_A,\ldots,N_O\}$ is 
given by identifying the remaining $N_{\alpha}^{\rm CFT}$ with the translation generators $\iota_1,\ldots,\iota_4$ on the set $\{N_A,\ldots,N_O\}$
\begin{equation}
\begin{alignedat}{5}
\label{dictionary}
	N_A &= \nc_{0011}\ , ~~ N_B &&= \nc_{0001}\ , ~~ N_C &&= \nc_{0010}\ , ~~ N_D &&= \nc_{1001}\ , ~~ N_E &&= \nc_{1010}\ ,\\
	N_F &= \nc_{1110}\ , ~~ N_G &&= \nc_{1101}\ , ~~ N_H &&= \nc_{1000}\ , ~~ N_I &&= \nc_{1011}\ , ~~ N_J &&= \nc_{0111}\ ,\\
	N_K &= \nc_{0100}\ , ~~ N_L &&= \nc_{0110}\ , ~~ N_M &&= \nc_{0101}\ , ~~ N_N &&= \nc_{1111}\ , ~~ N_O &&= \nc_{1100}\ .
\end{alignedat}
\end{equation}
The induced action of $A_8$ on this basis is shown to be given by
\begin{align}
\begin{split}
\label{a8rep}
	M(\alpha_1)&=(B,C)(D,L)(E,M)(F,G)(H,K)(I,J)\ ,\\
	M(\alpha_2)=M(\beta_2)&=(A,O)(B,K)(C,H)(D,L)(F,I)(G,J)\ ,\\
	M(\gamma_1)&=(A,C)(D,J)(E,M)(G,O)(H,L)(I,K)\ ,\\
	M(\gamma_2)&=(A,G)(C,O)(D,L)(E,I)(H,J)(K,M)\ ,\\
	M(\gamma_3)&=(A,H,I)(B,N,F)(C,J,M)(D,K,G)(E,O,L)\ ,\\
	M(\beta_1)&=(A,C,B)(D,N,L)(E,G,J)(F,M,I)(H,O,K)\ ,
\end{split}
\end{align}
whereas $\tg$ acts as $\pm 1$ on the ground states as in \eqref{z24rep}, and thus 
automatically stabilizes $N_A,\ldots,N_O$. Taken together this then defines a representation of $G_i$ on these $15$ twisted sector 
ground states. It is in fact a \emph{monomial representation}.\footnote{A
	monomial representation has the property that the representation matrices have the same non-zero patterns
	as permutation matrices. In our case, the non-zero entries are all $\pm 1$.}

Note  that $\tg$ acts trivially on $\nc_{0000}$, and
that there are no states invariant under it in any of the other 
15  sectors. This motivates an alternative definition
of $\Vp$ and its complement $\Vr$. Namely, we can define
$\Vr:= V^T$, that is the states that are invariant under
$T$; the projector onto $\Vr$ is thus 
\be
P_r := \frac{1}{16}\sum_{h \in \tg} h\ ,
\ee
while for the projector onto $\Vp$ we find
\be
P_\perp = 1 - P_r\ .
\ee
This gives indeed
\be
P_r \, \nc_\alpha = \delta_{\alpha,0}\, \nc_\alpha\ .
\ee

Let us now discuss the action on the Fock
space of oscillator modes.
The full space $\Vp$ is of the form 
\be
\Vp = \bigoplus_{X=A}^O V_X \ , 
\ee
where $V_X$ are the orbifold-invariant (left-moving) descendants of the ground state $N_X$. 
For each geometric subgroup, we can choose a basis of the oscillators
in the twisted sectors such that the action of $g\in G_i$ factorizes into an action on the 
oscillators, and an action on the ground states,
giving a tensor product representation 
\be\label{geomaction}
\rho^i= \rho^i_{\rm osc}\otimes \rho^i_{\rm gr}\ ,
\ee
where $\rho^i_{\rm gr}$ is the action on the ground
states defined at the beginning of this section.
In particular, for every $g$, $\rho^i$ defines a permutation
\be
\sigma^i_g \in S_{15}
\ee
such that
\be\label{rhoipermute}
\rho^i(g) V_X  \subset V_{\sigma_g(X)}\ .
\ee
$\sigma^i_g$ is of course the permutation generated by (\ref{a8rep}).

The representation $\rho^i_{\rm osc}$ on the other hand
can be embedded in the
${\rm SU}(2)$ representation (\ref{su2}) acting on the doublets
$\pl{i}$  and $\Xl{i}$. The representation
matrices of the generators for the $\mathbf{2}$ are given by
\begin{eqnarray}\label{repalpha1}
\alpha_1 =\gamma_1&=&\left(\begin{array}{cc}i&0\\0&-i\end{array}\right)\\
\label{repalpha2}
\alpha_2=\beta_2 =\gamma_2&=&\left(\begin{array}{cc}0&-1\\1&0\end{array}\right)\\
\beta_1&=&\left(\begin{array}{cc}\zeta&0\\0&\zeta^{-1}\end{array}\right)\\
\gamma_3&=&\left(\begin{array}{cc}\frac{-1+i}{2}&\frac{1-i}{2}\\ \frac{-1-i}{2}&-\frac{1+i}{2}\end{array}\right) \ ,
\label{repgamma3} 
\end{eqnarray}
where $\zeta = e^{2\pi i/3}$ --- see Proposition 3.4 in~\cite{Taormina:2013mda}.
For the $\bar{\mathbf{2}}$ we take, of course, the complex conjugate
matrices.
Note that this does not define
a representation of the $G_i$, but rather a projective
representation, whose phases are given by $\pm 1$. 
$\rho^i$ however is a proper representation due
to the fact that the $\Z_2$ orbifold ensures that only bilinear
modes of the oscillators survive, for which the sign is then always $+1$. This 
ties in with the fact that the orbifold should be crucial, as we do
not expect any interesting moonshine for the $\mathbb{T}^4$ theory, see however
\cite{Volpato:2014zla}.

In principle we could describe $\rho^i$ explicitly in terms of
the tensor product of
sums of tensor products of $\mathbf{2}$ and $\bar{\mathbf{2}}$ with 
the monomial representation of the ground states. In section~\ref{s:Results}
we will instead
simply give the (graded) characters of the $\rho^i$,
which is of course enough to uniquely identify them.

In summary, the action of the geometric group $G_i$ factorizes as in (\ref{geomaction}), where 
$\rho_i^{\rm osc}$ acts on the oscillator modes as in (\ref{repalpha1}) -- (\ref{repgamma3}), and $\rho^i_{\rm gr}$ is
a monomial matrix given by (\ref{z24rep})  and (\ref{a8rep}).
This is the analogue of the statement in \cite{Taormina:2013mda} that for the $\mathbf{45}$, for each geometric subgroup individually one can find a basis such that there is no `twist', \ie that the induced permutations $m_\tau^{(X,\tau(X))}$ are independent of the sector $X$. 

\subsection{Representations of $G$}\label{ss:Grep}

As we saw in Section~\ref{geomgroup}, by embedding the different symmetry groups into the permutation group $S_{24}$, 
there is a natural way of combining them into the octad group $G$. However, 
it is not immediately clear how to combine the actions of the geometric
groups into an action of the full octad group $G$. In particular, 
the fact that the geometric representations can be written as
tensor products (\ref{geomaction}) depended on a specific choice
of basis for each $G_i$. These choices of basis are in general not compatible
with each other.
What we can do is to try to translate the different group actions
by parallel transporting the BPS states between the three special points.
This
will, in general, lead to a non-trivial change of basis, i.e., the different group actions will be `twisted' relative to one another. 

More specifically, let us denote by $\rho^i_{p_i}$ the action of the group $G_i$ at the point $p_i$ in the moduli space where the Kummer surface
actually has the enhanced symmetry $G_i$. We can
then define the action $\rho^i_{p_j}$ of the group $G_i$
at another point $p_j$ by parallel transport, that is
\be\label{parallel}
\rho^i_{p_j} := \gamma_{ji}\, \rho^i_{p_i}\, \gamma_{ji}^{-1}\ ,
\ee
where $\gamma_{ji}$ is the matrix that implements the parallel
transport from $p_i$ to $p_j$. (Since we may connect the three different
points in moduli space within the Kummer moduli space, we do not expect
that this will mix the $15$ surviving twisted sectors with the rest, in particular,
the untwisted sector. Thus it makes sense to analyse this on the subspace
$\Vp$.) We expect $\gamma_{ij}$ to be block-diagonal with $15$ blocks; however,
the different blocks will not be in general identical --- this is what \cite{Taormina:2013mda}
call the `twist'. Using this definition, we can
then define a representation $\rho$ of the full octad group $G$.
We define it for instance at the point $p_0$  
by combining the $\rho^i_{p_0}$ for $i=0,1,2$, that is
\be
\rho(g) := \rho^i_{p_0}(g) \qquad {\rm if}\ g \in G_i\ .
\ee 
Let us now discuss what properties $\rho$ has to satisfy
so that this proposal is consistent.
First, clearly the restriction to the geometric groups
has to satisfy
\be
{\rm Res}_{G_i} \rho = \rho^i_{p_0} \cong \rho^i\ ,
\ee
since $\rho^i_{p_0}$ and $\rho^i$ only differ by
a change of basis. This explains why we want
to prove the second statement of Theorem~\ref{thm1}. Before
we get to this, let us first explain the origin of the induced representations,
see the first statement of Theorem~\ref{thm1}.

\subsubsection{Induced representations}

To start with, we concentrate on the ground 
states again. 
The discussion in section~\ref{ss:geoact} carries over directly,
so that we again get a monomial representation.
In fact we can describe this representation in a slightly different manner,
namely as an \emph{induced representation}. 

Let us chose one of the twisted sectors $X$, and consider the setwise stabilizer subgroup 
$\HA\subset G$ of $X$. By the orbit-stabilizer theorem, $\HA$ has index 15. 
Note that the conjugacy class of $\HA$ is independent of the choice of $X$, which
is why we will write $\HA$ rather than $\HA^X$. With $g_i$, $i=1,\ldots, 15$ a set of
representatives of $G/\HA$, we can then write the representation as
\be
\mathbf{15} = \bigoplus_{i=1}^{15}\C  g_i\ .
\ee
Since $g_i\in G/\HA$ there is a natural action from the left of $G$
which defines the representation $\rho(G)$.
In fact it is easy to check that this is precisely the $\mathbf{15}$
representation of the octad group $G$.

Next we want to describe the group action on a general element of $\Vp$. 
We 
could of course construct the full representation $\rho$
if we knew the transport matrices $\gamma_{ij}$.
In principle we could obtain them through conformal
perturbation theory. We have not attempted to do this,
and will instead just use some of their general properties.
In particular we do not expect parallel transport to
mix different twist sectors, but rather only transform
states within a given sector. This means that the transport matrices $\gamma_{ij}$ are block diagonal, so that the
representation matrices $\rho$ still have the same form as (\ref{rhoipermute}),
\be\label{rhopermute}
\rho(g) V_X \subset V_{\sigma_g(X)} \qquad X=A,B,\ldots,O\ ,
\ee
where $\sigma_g$ can again be obtained from (\ref{a8rep}).
Note however that because of the $\gamma_{ij}$, $\rho$ is
no longer simply a tensor product of a monomial representation
with another representation. However, (\ref{rhopermute}) does mean that
$\rho$ is a so-called \emph{imprimitive} representation, see, e.g., \cite{imprimitive_group}. The system
of imprimitivity is $V_A,\ldots, V_O$. It is straightforward to see that $G$ acts transitively on it. 
From a general theorem of representation theory, see again, e.g., \cite{imprimitive_group},
it follows that $\rho$ can be written as an induced representation
of the stabilizer subgroup $\HA$.

More explicitly, let $\rho^N$ be the representation
of $\HA$ on $V_N$. We can then write
the representation $\Vp$ as the induced representation
\be
\Vp = \textrm{Ind}_{\HA}^G\,  \rho^{N} = \bigoplus_{i=1}^{15}  g_i V_N\ .
\ee
Conversely,
note that the permutations $\sigma_g$ of the induced representation is
fixed by $\HA$ and the action of $G$ on the cosets $G/\HA$, and
is therefore independent of $\rho^N$. This means, in particular,
that representations induced from any representation $\rho^N$
of $\HA$ will be permuted in the correct way to be compatible
with our proposal. The prediction is thus that all
representations of $G$ that appear in $\Vp$ are in fact
induced from representations $\rho^N$ of the stabilizer group $\HA$.

The sector $N_N$ which we have chosen is mapped to itself by $\tg$ and the generators $\alpha_1$, 
$\alpha_2$, $\gamma_1$ and $\gamma_2$. The vector-space stabilizer 
Stab($N_N$), is given by 
$\HA:=((\z_2)^6\rtimes PSL(3,2))\rtimes\z_2$ and, as expected, is an index $15$ subgroup of $G$. It is 
generated by the $\tg$ generators together with 
\begin{align}
\begin{split}
	\eta_1&:=\alpha_1\cdot\gamma_1\cdot\alpha_1\ ,\\
	\eta_2&:=\gamma_3\cdot\alpha_2\cdot\alpha_1\cdot\gamma_1\cdot\beta_1\cdot\alpha_1\ ,\\
	\eta_3&:=\alpha_1\ .
\end{split}
\end{align}
If the proposal of symmetry surfing is correct, we thus expect
that all representations in $\Vp$ are induced from representations
of $\HA$. This explains the motivation behind the first statement
in Theorem~\ref{thm1}.

\section{Results}\label{s:Results}
\subsection{Induced representations}
In section~\ref{ss:Grep} we motivated the statements of Theorem~\ref{thm1}. 
Let us now prove them.
We first prove the first statement 
by checking that $\Vp$ is indeed an 
induced representation of $\HA$. 
As mentioned above, we define the projector $P_r$ as
\be\label{Prest}
P_r = \frac{1}{16}\sum_{h\in\tg}h\ ,
\ee
and $P_\perp$ as the projector on the orthogonal
complement
\be
P_\perp = 1 - P_r\ .
\ee
This allows us to define $\Vp$ and $\Vr$ as
\be
\Vp = P_\perp V\ , \qquad \Vr = P_r V\ .
\ee
Since $\tg$ is a normal subgroup of $G$,
$P_r$ commutes with the action of $G$. The
 irreducible representations of
$G$ thus fall into two sets, 
\be\label{Vperpdecomp}
\Vp : \qquad \cR_{\rm tw} \equiv \{\textbf{15}, \textbf{45}',\con{45}',\textbf{90},\textbf{105}_1,
\textbf{105}_2,\textbf{105}_3,\textbf{120},\textbf{210},\textbf{315}_1,\textbf{315}_2 \}\ ,
\ee
and
\be\label{Vrestdecomp}
\Vr: \qquad  \cR_0 \equiv \{\textbf{1},\textbf{7},\textbf{14},\textbf{20},\textbf{21},\textbf{21}',\con{21}',
\textbf{28},\textbf{35},\textbf{45},\con{45},\textbf{56},\textbf{64},\textbf{70}\}\ .
\ee
This is easily shown using the character table of $G$
in appendix~\ref{app:chartable}: all 15 
non-trivial elements of $\tg$ are in the 
conjugacy class $\CTwo$, so that indeed
\be
\Tr_{\Vr} P_r = \dim \Vr \qquad \textrm{for} \ \Vr \in \cR_0
\ee
and
\be
\Tr_{\Vp} P_r = \dim  \Vp - \frac{\dim \Vp}{15}\cdot 15 = 0
\qquad \textrm{for} \ \Vp \in \cR_{tw}\ .
\ee
We then check using GAP \cite{GAP4} that all representations of $G$ in $R_{\rm tw}$ are induced representations, which proves the statement.
Note that $R_0$ also contains two induced representations, namely $\textbf{45}$ and $\con{45}$; the complete 
set of induced representations is given in Table~\ref{tab1}.

\begin{table}[h]
\centering
\begin{tabular}{| >{$}c<{$} || >{$}c<{$} | >{$}c<{$} | >{$}c<{$} | >{$}c<{$} | >{$}c<{$} | >{$}c<{$} | >{$}c<{$} 
| >{$}c<{$} | >{$}c<{$} | >{$}c<{$} | >{$}c<{$} | >{$}c<{$} | >{$}c<{$} | }
\hline
G&\textbf{15}& \textbf{45}'&\con{45}' &\textbf{45}&\con{45}&\textbf{90}&\textbf{105}_1&\textbf{105}_2&\textbf{105}_3&\textbf{120}
&\textbf{210}&\textbf{315}_1&\textbf{315}_2\\	
\HA&\textbf{1}'&\textbf{3}&\con{3}&\textbf{3}'&\con{3}'&\textbf{6}'&\textbf{7}_4&\textbf{7}_2&\textbf{7}_5&\textbf{8}'&\textbf{14}_2&\textbf{21}_2&\textbf{21}_3\\
\hline
\end{tabular}
\caption{The $G$ representations that are induced from $\HA$, and the  $\HA$ representations from which
they are induced. For the character tables of the $\HA$ representations see table~\ref{t:charTableH} in appendix~\ref{app:chartable}.}
\label{tab1}
\end{table}

Since we know the decomposition of the elliptic genus in 
terms of \mg\ representations, and since the octad group $G$ is a subgroup  of \mg, $G\subset\mg$, 
we 
can unambiguously decompose all \mg\ representations in terms of $G$ representations. Then we divide these representations into the 
contributions coming from $\Vp$ and $\Vr$, respectively, using the division coming from $\cR_{\rm tw}$ and $\cR_{0}$. 
This analysis can be done once and for all for each \mg\ 
representation, and the result of this analysis is summarised in Table~\ref{tab:split}. 
With this knowledge at hand, we can now take the known decomposition of the elliptic genus in terms
of \mg\ representations, and deduce from it the decomposition in terms of $\Vp$ and $\Vr$; for the first
few levels this is done explicitly in Table~\ref{tab:splitting1}, see appendix~\ref{app:D}. 

\bgroup
\def\arraystretch{1.2}
\setlength\tabcolsep{4pt}
\begin{table}%
{\small
\begin{tabular}{| >{$}c<{$} || >{$}c<{$} | >{$}c<{$} |}
\hline
\mg&\vo&\vr\\\hline

\textbf{1}& \cdot&\textbf{1}\\\hline

\textbf{23}&\textbf{15}&\textbf{1}\oplus\textbf{7}\\\hline

\textbf{45}&\textbf{45}'& \cdot\\\hline

\con{45}&\con{45}'& \cdot\\\hline

\textbf{231}&\textbf{210}&\textbf{21}'\\\hline

\con{231}&\textbf{210}&\con{21}'\\\hline

\textbf{252}&\textbf{15}\oplus\textbf{90}\oplus\textbf{105}_3&\textbf{1}\oplus\textbf{7}\oplus\textbf{14}\oplus\textbf{20}\\\hline

\textbf{253}&\textbf{15}\oplus\textbf{105}_1\oplus\textbf{105}_3&\textbf{7}\oplus\textbf{21}\\\hline

\textbf{483}&\textbf{15}\oplus\textbf{90}\oplus\textbf{105}_3\oplus\textbf{210}&\textbf{1}\oplus\textbf{14}\oplus\textbf{20}\oplus\textbf{28}\\\hline

\textbf{770}&\textbf{105}_2\oplus\textbf{315}_1\oplus\textbf{315}_2&\textbf{35}\\\hline

\con{770}&\textbf{105}_2\oplus\textbf{315}_1\oplus\textbf{315}_2&\textbf{35}\\\hline

\textbf{990}& 2\cdot\textbf{315}_1\oplus\textbf{315}_2&\textbf{45}\\\hline

\con{990}& 2\cdot\textbf{315}_1\oplus\textbf{315}_2&\con{45}\\\hline

\textbf{1035}&\textbf{90}\oplus\textbf{105}_2\oplus\textbf{105}_3\oplus\textbf{120}\oplus\textbf{210}\oplus\textbf{315}_2&\textbf{14}\oplus\textbf{20}\oplus\textbf{56}\\\hline

\textbf{1035}'&\textbf{45}'\oplus 2\cdot\textbf{315}_1\oplus\textbf{315}_2&\textbf{45}\\\hline

\con{1035}'&\con{45}'\oplus 2\cdot\textbf{315}_1\oplus\textbf{315}_2&\con{45}\\\hline

\textbf{1265}&\textbf{15}\oplus 2\cdot\textbf{90}\oplus\textbf{105}_1\oplus\textbf{105}_3\oplus 2\cdot\textbf{210}\oplus\textbf{315}_2&\textbf{7}\oplus\textbf{20}\oplus\textbf{28}\oplus\textbf{70}\\\hline

\textbf{1771}&\textbf{105}_1\oplus 2\cdot\textbf{105}_2\oplus\textbf{105}_3\oplus 2\cdot\textbf{315}_1\oplus 2\cdot\textbf{315}_2&\textbf{21}\oplus 2\cdot\textbf{35}\\\hline

\multirow{2}{*}{\textbf{2024}}&\textbf{90}\oplus\textbf{105}_2\oplus\textbf{105}_3\oplus\textbf{120}&\multirow{2}{*}{$\textbf{14}\oplus\textbf{56}\oplus\textbf{64}$}\\
&\oplus~\textbf{210}\oplus 2\cdot\textbf{315}_1\oplus 2\cdot\textbf{315}_2&\\\hline

\multirow{2}{*}{\textbf{2277}}&\textbf{15}\oplus\textbf{90}\oplus\textbf{105}_1\oplus\textbf{105}_3\oplus\textbf{120}&\multirow{2}{*}{$\textbf{28}\oplus\textbf{64}\oplus\textbf{70}$}\\
&\oplus~2\cdot\textbf{210}\oplus 2\cdot\textbf{315}_1\oplus 2\cdot\textbf{315}_2&\\\hline

\multirow{2}{*}{\textbf{3312}}&\textbf{15}\oplus 2\cdot\textbf{90}\oplus\textbf{105}_1\oplus\textbf{105}_2\oplus 2\cdot\textbf{105}_3&\multirow{2}{*}{$\textbf{14}\oplus\textbf{20}\oplus\textbf{28}\oplus\textbf{56}\oplus\textbf{64}\oplus\textbf{70}$}\\
&\oplus~2\cdot\textbf{120}\oplus 3\cdot\textbf{210}\oplus 2\cdot\textbf{315}_1\oplus 3\cdot\textbf{315}_2&\\\hline

\multirow{2}{*}{\textbf{3520}}&\textbf{15}\oplus 3\cdot\textbf{90}\oplus 2\cdot\textbf{105}_1\oplus\textbf{105}_2\oplus 3\cdot\textbf{105}_3&\textbf{7}\oplus\textbf{14}\oplus\textbf{20}\oplus\textbf{21}\oplus\textbf{28}\\
&\oplus~\textbf{120}\oplus 3\cdot\textbf{210}\oplus  2\cdot\textbf{315}_1\oplus 3\cdot\textbf{315}_2&\oplus~\textbf{56}\oplus\textbf{64}\oplus\textbf{70}\\\hline

\multirow{2}{*}{\textbf{5313}}&3\cdot\textbf{90}\oplus\textbf{105}_1\oplus\textbf{105}_2\oplus 2\cdot\textbf{105}_3&\textbf{20}\oplus\textbf{21}'\oplus\con{21}'\oplus 2\cdot\textbf{56}\\
&\oplus~3\cdot\textbf{120}\oplus 5\cdot\textbf{210}\oplus 4\cdot\textbf{315}_1\oplus 5\cdot\textbf{315}_2&\oplus~\textbf{64}\oplus 2\cdot\textbf{70}\\\hline

\multirow{2}{*}{\textbf{5544}}& 2\cdot\textbf{45}'\oplus 2\cdot\con{45}'\oplus 2\cdot\textbf{105}_1\oplus 3\cdot\textbf{105}_2\oplus\textbf{105}_3&\multirow{2}{*}{$ 2\cdot\textbf{35}\oplus\textbf{45}\oplus\con{45}\oplus\textbf{64}\oplus\textbf{70}$}\\
&\oplus~2\cdot\textbf{120}\oplus 2\cdot\textbf{210}\oplus 6\cdot\textbf{315}_1\oplus 6\cdot\textbf{315}_2&\\\hline

\multirow{2}{*}{\textbf{5796}}& 2\cdot\textbf{45}'\oplus 2\cdot\con{45}'\oplus\textbf{90}\oplus\textbf{105}_1\oplus 2\cdot\textbf{105}_2\oplus\textbf{105}_3&\multirow{2}{*}{$\textbf{45}\oplus\con{45}\oplus 2\cdot\textbf{56}\oplus\textbf{64}\oplus\textbf{70}$}\\
&\oplus~3\cdot\textbf{120}\oplus 3\cdot\textbf{210}\oplus 6\cdot\textbf{315}_1\oplus 6\cdot\textbf{315}_2&\\\hline

\multirow{2}{*}{\textbf{10395}}&\textbf{45}'\oplus\con{45}'\oplus 2\cdot\textbf{90}\oplus 4\cdot\textbf{105}_1\oplus 3\cdot\textbf{105}_2\oplus 3\cdot\textbf{105}_3&\textbf{21}\oplus\textbf{21}'\oplus\con{21}'\oplus\textbf{28}\oplus\textbf{35}\oplus 2\cdot\textbf{45}\\
&\oplus ~3\cdot\textbf{120}\oplus 7\cdot\textbf{210}\oplus 11\cdot\textbf{315}_1\oplus 10\cdot\textbf{315}_2& \oplus~2\cdot\con{45}\oplus\textbf{56}\oplus 2\cdot\textbf{64}\oplus 2\cdot\textbf{70}\\\hline

\end{tabular}}
\caption{Decomposition of $\mg$ irreducible representations into Octad irreps and their splitting into the subspaces $\vo$ and $\vr$.}
\label{tab:split}
\end{table}}

\subsection{Geometric representations}
The first consistency
check is then that this reproduces correctly the graded dimensions of $\Vp_h$. In particular, the 
sums of dimensions that appear in the second column of Table~\ref{tab:splitting1} in appendix~\ref{app:D} 
must add up to $15 D_n$, see eq.~(\ref{ucoeff0}), and this indeed turns out to be correct.

In this section we prove the second statement of 
Theorem~\ref{thm1}, that is that the restriction of $\Vp_h$ to any of the three geometric subgroups $G_i$, $i=0,1,2$,  is isomorphic 
to the representations $\rho^{G_i}$,
$$
\mathrm{Res}_{G_i}\, \Vp_h \cong \rho_h^{G_i} \ .
$$
We will establish this by comparing characters.

In order to do this calculation efficiently, it is much easier to evaluate the full character
including the ${\cal N}=4$ descendants, rather than restrict to the ${\cal N}=4$ primary
states (that are counted by $\Vp_h$). Let us denote the full space (including descendants)
by $\Vpd$. For the purposes of the representation theory of $G$, $\Vpd_h$ is then
simply a direct sum of $\Vp_h$ and the lower lying $\Vp_h$, 
the multiplicities given by the number of descendants,
which we can read off from the known $\cN=4$ characters.
For the first few cases we have
\begin{eqnarray}
\Vpd_{1/2}&=&\Vp_{1/2}\\
\Vpd_1 &=& \Vp_{1/2} \oplus \Vp_1 \\
\Vpd_2 &=& 5\cdot \Vp_{1/2} \oplus 7\cdot\Vp_1 \oplus \Vp_2\\
\Vpd_3 &=& 14\cdot\Vp_{1/2} \oplus 25\cdot\Vp_1
\oplus 7\cdot \Vp_2\oplus \Vp_3\\
\Vpd_4 &=& 36\cdot \Vp_{1/2}\oplus 70\cdot \Vp_1\oplus 25\cdot \Vp_2\oplus 7\cdot \Vp_3\oplus\Vp_4 \ ,  
\end{eqnarray}
etc. From the triangular form of this, it is clear that if the
characters agree on the $\Vpd_h$, they will also
agree on the $\Vp_h$, and vice versa.
The only slight subtlety here is that we also need to know the representation of $\Vp_{1/2}$,
which contributes to the massless ${\cal N}=4$ representations, i.e., to the coefficient of 
$20$ in (\ref{1.1}). As was explained in section~\ref{ss:geoact}, this representation
is simply 
\be
\Vp_{1/2}=\Vpd_{1/2} = \bf{15}\oplus\bf{15}\ .
\ee
Note that this is compatible with ${\bf 20} = {\bf 23} - 3 \cdot {\bf 1}$, using the fact that 
the $\Vp$ part of ${\bf 23}$ is precisely ${\bf 15}$. 

We define the specialized graded character as
\be
\phi^\perp_g (\tau):= \Tr_{\Vpd}\Bigl( g \, q^{L_0-1/4} \Bigr)\ .
\ee
If $g$ is a geometric element, \ie if $g$ is in
one of the geometric groups $G_i$, we can evaluate
$\phi^\perp_g(\tau)$, since we know $\Vpd$ explicitly,
and in this case we also know the action of $g$.
For the details of this computation, see appendix~\ref{app:geo}.

As expected, the result shows that for all geometric
$g$, $\phi^\perp_g$ 
only depends on the $G$-conjugacy class of $g$.
In fact all geometric elements fall into
$5$ out of the $25$ conjugacy classes of $G$, which are denoted by $\COne$,  $\CTwo$, $\CSeven$, $\CEleven$, $\CTwelve$, 
with $1\in \COne$. The result is
\be\label{res1}
\phi^\perp_g(\tau)= \left\{\begin{array}{cc} 
15 \hat\phi_1(\tau) &: g = e\\
-\hat\phi_1(\tau) &: g \in \CTwo\\
0&: g\in \CSeven\\
3\hat\phi_i(\tau)&: g\in \CEleven\\
-\hat\phi_i(\tau)&: g\in \CTwelve
\end{array}	
\right.
\ee
where
\be
\hat\phi_1(\tau)= T_{l=0}(\tau,0)= \frac12
\frac{\vartheta_2(\tau)^2}{\vartheta_4(\tau)^2}\ 
\ee
and
\be
\hat\phi_i(\tau)= \frac{\vartheta_2(2\tau)}{\vartheta_3(2\tau)}\ .
\ee
Knowing $\hat\phi(\tau)$ fixes the $\rho_h^{G_i}$.
In principle we could of course construct them explicitly
from the Fock space of the torus orbifold, but we will
be satisfied by computing their characters.

It remains to compute the characters of the left-hand side
of (\ref{Thm1}). To this end we use the fact that the
\mg twining genera are known; they are defined as
\be
\chi_g(\tau)= \sum_n \chi_n(g)q^{n-1/4}\ ,
\ee
where $\chi_n(g)$ is the character of the
\mg\ representation $W_n$. To obtain the
analogue for $\Vpd$, we use the fact that
the projector $P_\perp$ can be written as
(\ref{Prest}). This leads us to define 
\be\label{chiperp}
\chi^\perp_g(\tau):= 
\chi_g(\tau)- \frac{1}{16}\sum_{h\in\tg}\chi_{hg}(\tau)\ ,
\ee
which we can determine from the knowledge of the twining genera
$\chi_{\tilde{g}}(\tau)$, which were all obtained in
\cite{Cheng:2010pq,Gaberdiel:2010ch,Gaberdiel:2010ca,Eguchi:2010fg}.
We will follow the conventions and notation of \cite{Gaberdiel:2010ch}, except
for an overall factor of $-2$ in the definition of the $\chi_g(\tau)$. 
We will denote the conjugacy classes of \mg\ by uppercase
letters, whereas we denote conjugacy classes of $G$ by lowercase
letters, taking their labeling from the character table~\ref{t:charTableG}.
They are then related by 
\be
\COne = 1{\rm A} \ ,  \quad \CTwo \subset 2{\rm A}\ ,\quad \CSeven \subset 3{\rm A} \ ,
\quad \CEleven \subset 2{\rm A}\ ,\quad \CTwelve \subset 4{\rm B} \ .
\ee
We can then evaluate (\ref{chiperp})
by using the expressions
\bea
\chi_{{\rm 1A}}(\tau)&=&-8\frac{\vartheta_4(\tau)^2}{\vartheta_2(\tau)^2}+8\frac{\vartheta_2(\tau)^2}{\vartheta_4(\tau)^2} \label{chi1A}\ , \\
\chi_{{\rm 2A}}(\tau)&=&-8\frac{\vartheta_4(\tau)^2}{\vartheta_2(\tau)^2}\ , \\
\chi_{{\rm 3A}}(\tau)&=& -2\left(\frac{\eta(\tau)}{\eta(3\tau)}\right)^3
\frac{\vartheta_3(3\tau)}{\vartheta_3(\tau)} \\
\chi_{{\rm 4B}}(\tau)&=&-4\frac{\vartheta_3(2\tau)}{\vartheta_2(2\tau)}\ .
\eea
For $g=e$ we get 
\be
\chi^\perp_g(\tau) = \frac{15}{16}\chi_{\rm 1A}(\tau) - \frac{15}{16}\chi_{\rm 2A}(\tau)= 15T(\tau)\ ,
\ee
while for $g\in \CTwo$ we find 
\be
\chi^\perp_g(\tau) = \frac{1}{16} \bigl(\chi_{\rm 2A}(\tau)-\chi_{\rm 1A}(\tau)\bigr)= -T(\tau) \ .
\ee
For $\CSeven$, we pick the representative 
$\gamma_3 \in \CSeven$. We find that $h \gamma_3 \in 3A$
for all $h\in (\z_2)^4$, 
so that
\be
\chi^\perp_g(\tau) = \chi_{\rm 3A}(\tau) - \chi_{\rm 3A}(\tau) = 0\ .
\ee
For $\CEleven$, we pick $\gamma_1\in \CEleven$. We find that 3 of the nontrivial
elements of $(\z_2)^4$ are mapped to $2{\rm A}$, and the other 12 to $4{\rm B}$, 
giving in total
\be
\chi^\perp_g(\tau) 
=\frac{3}{4} \bigl(\chi_{\rm 2A}(\tau)-\chi_{\rm 4B}(\tau)\bigr) \ .
\ee 
For $\CTwelve$, picking $\iota_1 \gamma_1 \in \CTwelve$ we find that 4 of the nontrivial
elements of $(\z_2)^4$ are mapped to ${\rm 2A}$, and the other 11 to ${\rm 4B}$, giving
\be
\chi^\perp_g(\tau) 
=-\frac{1}{4}\bigl(\chi_{\rm 2A}(\tau)-\chi_{\rm 4B}(\tau)\bigr) \ .
\ee
Using the identity
\be
-\frac{1}{4}\chi_{\rm 4B}(\tau)+\frac{1}{4}\chi_{\rm 2A}(\tau)=
\frac{\vartheta_3(2\tau)}{\vartheta_2(2\tau)}
-2\frac{\vartheta_4(\tau)^2}{\vartheta_2(\tau)^2}
= \frac{\vartheta_2(2\tau)}{\vartheta_3(2\tau)}
\ ,
\ee
which follows directly from \cite{GR} eqs.~(8.199) 3.\ and 6., we find in all cases
\be
\phi^\perp_g(\tau)=\chi^\perp_g(\tau)\ .
\ee

Finally let us show that the geometric action on $\Vr$
is compatible with the \mg\ representation theory.
In appendix~\ref{app:geo}, see in particular section~\ref{app:geocharsrest}, we compute the geometric characters 
$\phi^r_g(\tau)$ for $\Vr$. We find
\be
\phi_g^{r}(\tau) = \left\{ \begin{array}{cc}   -8\frac{\vartheta_4(\tau)^2}{\vartheta_2(\tau)^2} + 
	\hat\phi_1(\tau)
	\qquad & g \in \COne,\CTwo\\
	\phi^{\rm ut}_\zeta(\tau)  + 
	\hat\phi_\zeta(\tau) \qquad & g\in \CSeven\\
	-4 \frac{\vartheta_3(2\tau)}{\vartheta_2(2\tau)} + \hat{\phi}_i (\tau)  \qquad & g\in \CEleven, \CTwelve 
\end{array}\right .
\ee
Together with (\ref{res1}) and (\ref{hatphisimple}) it then follows that 
\be
\phi_g^r(\tau)+\phi_g^\perp(\tau) =   \chi_g(\tau)\ ,
\ee
at least for $g\in \COne,\CTwo$ and $g\in \CEleven, \CTwelve$; for $g\in \CSeven$ we have only verified this identity to high order --- 
but because of the modular properties of the functions in question, this is then also sufficient to show the result. Taken together this therefore 
proves Theorem~\ref{thm1}.  $\square$

\section{Conclusions}\label{s:conclusions}

In this paper we have subjected the symmetry surfing idea of Taormina and Wendland  \cite{Taormina:2013mda,Taormina:2013jza} 
to a stringent consistency check. In particular, their idea predicts that the surviving contribution from the twisted sector (after
performing some deformation to move off the orbifold moduli space) should furnish a representation of the octad group.
The octad group is not the symmetry group of an individual K3 sigma model, but arises upon putting the symmetries
of three special sigma models together. The fact that we find very convincing evidence for the presence of this octad
symmetry is therefore a non-trivial confirmation of their philosophy.

The octad approach also seems promising from another viewpoint. As we have explained in the paper, 
there is a simple division of octad representations according to whether they arise in the orthogonal part of the twisted 
sector $\Vp$, or in the rest, $\Vr$, see eqs.~(\ref{Vperpdecomp}) and (\ref{Vrestdecomp}). Since the representations
in ${\cal R}_0$ are precisely those representations that are invariant under $\tg$, while ${\cal R}_{\rm tw}$ contains the remaining
representations,  
the tensor products of representations in ${\cal R}_{0}$ close among themselves, while those involving a representation
in ${\cal R}_0$ and one in ${\cal R}_{\rm tw}$ lie in ${\cal R}_{\rm tw}$, i.e., 
\be\label{subalgebra}
{\cal R}_0 \otimes {\cal R}_0 \subseteq {\cal R}_0 \ , \qquad {\cal R}_0 \otimes {\cal R}_{\rm tw} \subseteq {\cal R}_{\rm tw} \ ;
\ee
incidentally, this property can also be checked directly. 
(The tensor products of representations in ${\cal R}_{\rm tw}$ with themselves lead to representations both in 
${\cal R}_0$ and ${\cal R}_{\rm tw}$.) 

This observation is nicely compatible with the idea of an underlying algebra structure. One interpretation is that $\Vr$ is 
associated to an $\tg$ invariant subspace of a VOA-like structure, and that $\Vp$ is a $\z_2^4$-twisted module; this is 
also compatible with the $S$ modular transformation properties of 
$\phi_g^r(\tau)$ and $\phi_g^\perp(\tau)$. It would be very interesting to explore this idea further.

\bigskip
\centerline{\bf{Acknowledgements}}
\bigskip

This paper is based on the Master thesis of one of us (H.P.). 
We thank Jeff Harvey and Greg Moore for initial collaboration on this project.
It is a pleasure to thank Simeon Hellerman, Geoff Mason, Anne Taormina, Roberto Volpato, Katrin Wendland for helpful discussions.
CAK thanks the Harvard University High Energy Theory
Group for hospitality. 
CAK is supported by the 
Swiss National Science Foundation through the NCCR SwissMAP. The research of MRG is also
supported partly by the NCCR SwissMAP, funded by the Swiss National Science Foundation.

\appendix

\section{The $\cN=4$ algebra at the orbifold point}
\label{app:N4}
We denote the complex fermions and complex bosons of $\mathbb{T}^4$
as 
\be
\pl{i} \ , \quad \pbl{i}  \ , \qquad  \Xl{i}\ , \quad \Xbl{i}  \ ,
\ee
where $i=1,2$. In what follows we will only discuss the left-movers.
The right-movers, which we denote by tilde, are completely analogous.
The modes of the left-moving fields obey the standard (anti-)commutation relations
\begin{align}
\label{commutation}
\begin{split}
	[\partial X^{i}_m, \bar{\partial X}^{j}_n]&=m\, \delta^{ij}\, \delta_{m,-n}\ ,\\
	\left\{\Psi_r^{i},\bar{\Psi}_s^{j}\right\}&=\delta^{ij}\, \delta_{r,-s}\ .
\end{split}
\end{align}
Sometimes we will also use uppercase indices to
write $\pl{I}, I=1,2,3,4$ with the understanding that $\pl{3}:=\pbl{1}$ etc.
In terms of these free fields, the $\hat{\mathfrak{su}}(2)_{R}$ $R$-currents
of the $\cN=4$ superconformal algebra are given by
\begin{eqnarray}\label{Jdef}
J^{+,3} & = & \frac{1}{2i} \Bigl(  \pl{1} \pbl{1} + \pl{2} \pbl{2} \Bigr) 
= \frac{1}{2i} \delta_{ij}\pl{i}\pbl{j}\ , \\[2pt]
J^{++} & = & - \pl{1} \pl{2} = -\frac{1}{2}\epsilon_{ij}\pl{i}\pl{j}\ ,\\[2pt]
J^{+-} & = & \pbl{1} \pbl{2} =\frac{1}{2}\epsilon_{ij}\pbl{i}\pbl{j} \ , 
\end{eqnarray}
where we work in the usual Cartan-Weyl basis for $\mathfrak{su}(2)$. 
The $\cN=4$ supercurrents are then given by
\begin{eqnarray}\label{Gdef}
G^{+} = \delta_{ij}\pl{i}\Xbl{j}, && G^{-}= \epsilon_{ij}\pbl{i}\Xbl{j}\ ,\\
G^{'+} = \epsilon_{ij}\pl{i} \Xl{j}, && G^{'-} = \delta_{ij}\pbl{i}\Xl{j}\ .\label{Gdef2}
\end{eqnarray}
The pairs $(G^+,G^-)$ and $(G^{'+}, G^{'-})$ form indeed doublets under
the zero modes of $\hat{\mathfrak{su}}(2)_{R}$.

The symmetry group for the torus orbifold  is actually bigger. In particular 
there is an additional ${\rm SU}(2)$ symmetry which commutes with the $\cN=4$ superconformal 
generators. The corresponding currents are given as 
 \begin{eqnarray}
J^{-,3} & = & \frac{1}{2i} \Bigl( - \pl{1} \pbl{1} + \pl{2} \pbl{2} \Bigr)\ , 
 \\[2pt]
J^{-+} & = & - \pl{1} \pbl{2} \label{J-}\ , \\[2pt]
J^{--} & = & \pbl{1} \pl{2} \ . 
\end{eqnarray}
With respect to this symmetry $\Psi^i$ and $\partial X^i$ then transform in the $\bf{2}$, 
while $\bar\Psi$ and $\partial \bar X$ transform in the $\bar{\bf{2}}$ (which is equivalent). 
In particular, the ${\cal N}=4$ superconformal fields $J$ and $G^\pm$ are invariant under these
${\rm SU}(2)$ transformations.

\section{Elliptic genus and characters}
\label{app:ellgenus}

\subsection{The untwisted sector}
\label{sec:untwist}
As a partial partition function, the elliptic genus in the R$\rt{\rm R}$ sector counts left-moving states with statistics 
signs due to the $(-1)^F$ factor in the trace. However, after spectral flow to the left-moving NS sector, we choose not to 
insert $(-1)^F$ so that all states are counted with the same sign. Furthermore, we choose the convention that 
positive charges are counted by 
positive powers of $y$, whereas negative charges give powers of $y^{-1}$. It is well known that bosonic states created 
by a single bosonic oscillator $a_n$ from the vacuum are counted by the infinite product $\prod_n (1-q^n)^{-1}$. Similarly, the states 
of a single positively charged fermion (with half-integer modes) are counted by the product $\prod_n(1+q^{n-\frac{1}{2}}y)$, and, 
accordingly, a negatively charged fermion results in the product $\prod_n(1+q^{n-\frac{1}{2}}y^{-1})$. In our K3 sigma model, we have 
four bosonic fields $\partial X^I$ together with two positively and two negatively charged fermions $\Psi^i$ and $\bar{\Psi}^i$, respectively. 
Putting  everything together, this then gives rise to a partition function of the form
\begin{equation}
\label{noorbifold}
	\prod_{n=1}^\infty\frac{(1+q^{n-\frac{1}{2}}y)^2(1+q^{n-\frac{1}{2}}y^{-1})^2}{(1-q^n)^4}\ .
\end{equation}
However, we have neglected the $\z_2$ orbifold projection so far. For the orbifold even ground state with right-moving $l=\frac{1}{2}$, 
only the states with an even number of modes contribute since all other states are projected out. In terms of partition functions, this 
can be done by adding a second term to the contribution~\eqref{noorbifold}. If we revers the sign in front of the $q$-monomials 
in the above expression, states with an odd number of modes will pick up a minus sign and therefore they cancel when both terms are 
added together. This yields the untwisted $l=\frac{1}{2}$ partition function
\begin{align}
\begin{split}
\label{ul12}
U_{l=\frac{1}{2}}(\tau,z) &= \frac{1}{2}q^{-\frac{1}{4}}\left(\prod_{n=1}^\infty\frac{(1+q^{n-\frac{1}{2}}y)^2(1+q^{n-\frac{1}{2}}y^{-1})^2}{(1-q^n)^4} 
+ \prod_{n=1}^\infty\frac{(1-q^{n-\frac{1}{2}}y)^2(1-q^{n-\frac{1}{2}}y^{-1})^2}{(1+q^n)^4} \right)\\
&= q^{-\frac{1}{4}}\left[1 + q \left(y^2+4+\frac{1}{y^2}\right) + q^{\frac{3}{2}}\left(8y+\frac{8}{y}\right)+q^2\left(4y^2+19+\frac{4}{y^2}\right) + \lo(q^{\frac{5}{2}})\right]\ ,
\end{split}
\end{align}
where the shift by $q^{-\frac{1}{4}}$ is due to the NS-vacuum, coming from the $q^{L_0-\frac{1}{4}}$ insertion in the definition of 
$\phi_{\rm K3}^{\rm NS}$, see \eqref{nsgen}.

On the other hand, the ground state with right-moving $l=0$ is orbifold odd, therefore only states with an odd number of modes 
created from this vacuum contribute to the elliptic genus. In that case, we have to cancel the even combinations and we find the 
untwisted $l=0$ partition function to take the form
\begin{align}
\begin{split}
\label{ul0}
	U_{l=0}(\tau,z) &= q^{-\frac{1}{4}}\left(\prod_{n=1}^\infty\frac{(1+q^{n-\frac{1}{2}}y)^2(1+q^{n-\frac{1}{2}}y^{-1})^2}{(1-q^n)^4} - \prod_{n=1}^\infty\frac{(1-q^{n-\frac{1}{2}}y)^2(1-q^{n-\frac{1}{2}}y^{-1})^2}{(1+q^n)^4}\right)\\
	&=q^{-\frac{1}{4}}\left[\sqrt{q}\left(4y+\frac{4}{y}\right)+8q+q^\frac{3}{2}\left(8y+\frac{8}{y}\right)+q^2\left(8y^2+40+\frac{8}{y^2}\right)+\lo(q^{\frac{5}{2}})\right],
\end{split}
\end{align}
where the factor of two from~\eqref{untwvacuum} is already included in this expression.\\ 
In order to find the number of primary fields in each sector, we decompose the given partition functions into $\cN=4$ characters. One finds
\begin{align}
\begin{split}
\label{udecomp}
		U_{l=\frac{1}{2}}(\tau,z) &= ch_{0,l=0}^{\rm NS}(\tau,z) + \sum_{n=1}^\infty{ B_n~ch_{h=n,l=0}^{\rm NS}(\tau,z)}\ ,\\
		U_{l=0}(\tau,z) &= 	4~ch_{0,l=\frac{1}{2}}^{\rm NS}(\tau,z) + \sum_{n=1}^\infty{ C_n~ch_{h=n,l=0}^{\rm NS}(\tau,z)}\ ,
\end{split}
\end{align}
where the coefficients $B_n$ and $C_n$ count the number of primary fields in each sector. They can be computed by a series expansion 
of the partition functions, using the explicit formulae for the ${\cal N}=4$ characters from \cite{Eguchi:1987wf,Eguchi:1988af}, 
and one finds
\begin{align}
\begin{split}
\label{ucoeff}
		B_n &= \{3,~1,~18,~15,~68,~89,~249,~358,~799,~1236,\ldots \}\ ,\\
		C_n &= \{0,~16,~8,~72,~80,~264,~360,~904,~1360,~2808,\ldots \}\ .
\end{split}
\end{align}

\subsection{The twisted sector}
\label{sec:twist}
In the twisted sector, the states are localized at fixed points of the $\mathbb{T}^4/\z_2$ orbifold. There are $2^4$ fixed points and thus we 
have 16 corresponding ground states which we denote by $\ket{\alpha}$, $\alpha\in\F_2^4$.
Since the orbifold projection exchanges the periodicity condition of fields, there are no fermionic zero modes in the twisted 
R$\rt{\rm R}$ sector. Therefore, the vacuum state is orbifold even and has quantum numbers $h=\rt{h}=\frac{1}{4}$ and $l=\rt{l}=0$.

However, we work in the NS$\rt{\rm R}$ sector, where due to spectral flow on the left-moving part one has integer fermionic modes 
and half-integer bosonic modes. This reintroduces left-moving fermionic zero modes such that the twisted ground states are given by
\begin{equation}
\label{twvacuum}
\ket{\alpha} := (\ket{l=\tfrac{1}{2}}\hspace{-5pt}\left.\right._{\rm NS}^\alpha\otimes\ket{l=0}_{\rt{\rm R}}^\alpha) ~\oplus~ 
2\cdot(\ket{l=0}_{\rm NS}^\alpha\otimes\ket{l=0}_{\rt{\rm R}}^\alpha)\ ,
\end{equation}
where the label $\alpha$ denotes one of the 16 fixed points. In order to account for these ground states, we simply 
multiply the partition function by a factor of $(y+2+y^{-1})$. Note that the left term of~\eqref{twvacuum} is orbifold 
even and the right one orbifold odd. As in the untwisted sector, we account for the orbifold projection by subtracting respectively 
adding a corresponding term in the partition function. The partition function for one 
twisted ground state $\ket{\alpha}$ is then given by
\begin{align}
\label{tw}
	T_{l=0}^\alpha(\tau,z) &= \frac{1}{2}q^{-\frac{1}{4}}\left( (y+2+y^{-1})~ q^{\frac{1}{2}} 
	\prod_{n=1}^\infty\frac{(1+q^{n}y)^2(1+q^n y^{-1})^2}{(1-q^{n-\frac{1}{2}})^4} -\right.\nonumber\\
	&\left.\qquad\qquad~~ -(-y+2-y^{-1})~ q^{\frac{1}{2}} \prod_{n=1}^\infty\frac{(1-q^{n}y)^2(1-q^n y^{-1})^2}{(1+q^{n-\frac{1}{2}})^4} \right)\\
	&=q^{-\frac{1}{4}}\left[\sqrt{q}\left(y+\frac{1}{y}\right)+8q+q^\frac{3}{2}\left(14y+\frac{14}{y}\right)
	+q^2\left(8y^2+64+\frac{8}{y^2}\right)+\lo(q^{\frac{5}{2}})\right]\ .\nonumber
\end{align}
Similarly, we can decompose $T_{l=0}^\alpha(\tau,z)$ into $\cN=4$ characters \cite{Eguchi:1987wf,Eguchi:1988af} according to
\begin{equation}
\label{twdecomp}
		T_{l=0}^\alpha(\tau,z) = 	ch_{0,l=\frac{1}{2}}^{\rm NS}(\tau,z) + \sum_{n=1}^\infty{ D_n~ch_{h=n,l=0}^{\rm NS}(\tau,z)}\ ,
\end{equation}
where the coefficients $D_n$ give the number of twisted primary fields at level $h=n$, namely
\begin{equation}
\label{twcoeff}
		D_n = \{6,~28,~98,~282,~728,~1734,~3864,~8182,~16618,\ldots\}\ .
\end{equation}

\subsection{Elliptic genus}

Finally, let us examine how the untwisted and twisted partition functions contribute to the elliptic genus. According to 
equation~\eqref{together} the untwisted and twisted sectors, expressed in terms of Jacobi theta functions, contribute as
\begin{align}
\begin{split}
\phi^{\rm NS}_{\rm untw}(\tau,z) =-2~U_{l=\frac{1}{2}}(\tau,z) + U_{l=0}(\tau,z) = -8\left(\frac{\vartheta_4(\tau,z)}{\vartheta_2(\tau,0)}\right)^2\ ,\\
\phi^{\rm NS}_{\rm tw}(\tau,z) =16~T^\alpha_{l=0}(\tau,z)=8\left(\frac{\vartheta_2(\tau,z)}{\vartheta_4(\tau,0)}\right)^2-8\left(\frac{\vartheta_1(\tau,z)}{\vartheta_3(\tau,0)}\right)^2\ ,
\end{split}
\end{align}
such that their sum $\phi^{\rm NS}_{\rm untw}(\tau,z)+\phi^{\rm NS}_{\rm tw}(\tau,z)$ correctly reproduces~\eqref{nsgenus}.

\section{Geometric Characters}\label{app:geo}

\subsection{Geometric Characters}\label{app:geochars}

In this appendix we compute the geometric characters.
For simplicity we will set $z=0$. We do not lose any information
that way, since the charges of states are essentially fixed
by the $N=4$ representation theory.
We therefor need to determine
\be
\phi_g^{\rm tw}(\tau)= \Tr_{\Vpd} \Bigl(g\, q^{L_0-1/4} \Bigr) 
\ee
and
\be
\phi_g^{\rm r}(\tau)= \Tr_{\Vrd} \Bigl(g\, q^{L_0-1/4} \Bigr) 
\ee
for all geometric elements $g$. As usual, it is enough
to do this for one representative per conjugacy class
of a geometric group.

An element $g$ acts as a permutation (\ref{a8rep}) on the twisted
sectors, and as an element of $SU(2)$ (\ref{repalpha1}) -- (\ref{repgamma3})
on the oscillator modes. Since we are computing the character,
the result only depends on the eigenvalues $\lambda$ and $\bar\lambda$
of the
$2\times2$ matrix $M(g)$, and the trace of the monomial representation,
\ie the number of fixed points with signs
of the permutation.

If $g$ has eigenvalues $\lambda = e^{2\pi i u}$ and $\bar\lambda = e^{-2\pi i u}$, the character
of its action on a single twisted sector is
\begin{align}\label{geotwist}
\hat\phi_\lambda(\tau)  & = \frac{q^{\frac{1}{4}}}{2} \Bigl( (2+\lambda +\bar{\lambda} ) 
\prod_{n=1}^\infty\frac{(1+\lambda q^{n})^2
	(1+\bar\lambda q^{n})^2}{(1-\lambda q^{n-\frac{1}{2}})^2(1-\bar\lambda q^{n-\frac{1}{2}})^2} \\
& \qquad +(2-\lambda -\bar{\lambda})
\prod_{n=1}^\infty\frac{(1-\lambda q^{n})^2(1-\bar\lambda q^{n})^2}
{(1+\lambda q^{n-\frac{1}{2}})^2(1+\bar\lambda q^{n-\frac{1}{2}})^2} \Bigr)
\\
&  = \frac{1}{2}\left(\frac{\vartheta_2(\tau,u)}{\vartheta_4(\tau,u)}\right)^2
+\frac{1}{2}\left(\frac{\vartheta_1(\tau,u)}{\vartheta_3(\tau,u)}\right)^2\ .
\end{align}
The form of the products follows from the action of $g$ on the oscillators.
The prefactors come from the fact that the left-moving ground states
$\Psi_0^1\ket{0}_{\rm NS}$ and $\Psi_0^2\ket{0}_{\rm NS}$ are
orbifold odd and have eigenvalues $\lambda$ and $\bar \lambda$,
whereas $\ket{0}_{\rm NS}$ and $\Psi_0^1\Psi_0^2\ket{0}_{\rm NS}$ are
orbifold even and invariant under $g$. The right-moving Ramond
ground state $\ket{0}_{\rm  \tilde R}$ is also invariant under $g$.
The total result in the twisted sector is thus
\be\label{phihatu}
\phi_g^{\rm tw}(\tau) = N_{\rm fp}(g)\,  \hat\phi_\lambda(\tau)\ ,
\ee
where $N_{\rm fp}(g)$ is the number of fixed points with signs.
In what follows, the eigenvalues $\lambda$ will always be 
$1,i$ or $\zeta=e^{2\pi i/3}$. For the first two cases we have
\be\label{hatphisimple}
\hat\phi_1(\tau) =  \frac{1}{2}\left(\frac{\vartheta_2(\tau)}{\vartheta_4(\tau)}\right)^2 = T_{l=0}(\tau,0)\ , \qquad 
\hat\phi_i(\tau) 
= \frac{\vartheta_2(2\tau)}{\vartheta_3(2\tau)}\ ,
\ee
while we have not managed to find a simple formula for the case $\lambda=\zeta = e^{2\pi i/3}$.

In the untwisted sector, the permutation part of the action
acts trivially. The character for the states in the right-moving
$l=\frac12$ representation is
\begin{align}
\phi^{l=\frac12}_\lambda(\tau) & =
\frac{q^{-1/4}}{2} \left( \prod _{n=1}^{\infty} \frac{(1+\lambda  q^{n-\frac{1}{2}})^2 (1+\bar{\lambda}q^{n-\frac{1}{2}})^2}{(1-\lambda  q^n)^2 (1-\bar{\lambda}q^n)^2}
+\prod _{n=1}^{\infty} \frac{(1-\lambda  q^{n-\frac{1}{2}})^2 (1-\bar{\lambda}q^{n-\frac{1}{2}})^2}{(1+\lambda  q^n)^2 (1+\bar{\lambda}q^n)^2}
\right)\\
& =2 \sin(\pi u)^2\frac{\vartheta_3(\tau,u)^2}{\vartheta_1(\tau,u)^2}
+ 2\cos(\pi u)^2\frac{\vartheta_4(\tau,u)^2}{\vartheta_2(\tau,u)^2} \ .
\end{align}
Note that from (\ref{rvacua}) the right-moving $l=\frac12$
representation in the Ramond sector is invariant under $g$ and is orbifold even. For
the right-moving $l=0$ representation we get
\begin{align}
\phi^{l=0}_\lambda(\tau) & =
\frac{\lambda+\bar{\lambda}}{2}q^{-1/4} \left( \prod _{n=1}^{\infty} \frac{(1+\lambda  q^{n-\frac{1}{2}})^2 (1+\bar{\lambda}q^{n-\frac{1}{2}})^2}{(1-\lambda  q^n)^2 (1-\bar{\lambda}q^n)^2}
-\prod _{n=1}^{\infty} \frac{(1-\lambda  q^{n-\frac{1}{2}})^2 (1-\bar{\lambda}q^{n-\frac{1}{2}})^2}{(1+\lambda  q^n)^2 (1+\bar{\lambda}q^n)^2}
\right)\\
& = 4\cos(2\pi u)\left( \sin(\pi u)^2\frac{\vartheta_3(\tau,u)^2}{\vartheta_1(\tau,u)^2}
- \cos(\pi u)^2\frac{\vartheta_4(\tau,u)^2}{\vartheta_2(\tau,u)^2}
\right) \ .
\end{align}
The prefactor comes from the eigenvalues of the right-moving
Ramond ground states $\Psi_0^1\ket{0}_{ \rm \tilde R}$
and $\Psi_0^2\ket{0}_{\rm \tilde R}$.
The untwisted contribution is thus
\be\label{phiut}
\phi^{\rm ut}_\lambda(\tau)=-2\phi_\lambda^{l=\frac12}(\tau)+\phi^{l=0}_\lambda(\tau)
= -8\left( 
\sin^4 \pi u \frac{\vartheta_3(\tau,u)^2}{\vartheta_1(\tau,u)^2}
+ \cos^4\pi u \frac{\vartheta_4(\tau,u)^2}{\vartheta_2(\tau,u)^2}
\right)\ ,
\ee
and in total we therefore get
\be\label{phir}
\phi^{\rm r}_g(\tau)=  \phi^{\rm ut}_\lambda(\tau)+
\hat\phi_\lambda(\tau)\ ,
\ee
where we included the contribution of the $G$-invariant
twisted sector. Again, there are simple formulae for the cases
\be
 \phi^{\rm ut}_1(\tau)=  -8\frac{\vartheta_4(\tau)^2}{\vartheta_2(\tau)^2} \ , 
\qquad \phi^{\rm ut}_i(\tau) = -4 \frac{\vartheta_3(2\tau)}{\vartheta_2(2\tau)} \ ,
\ee
while the expression for $\lambda=\zeta = e^{2\pi i/3}$ is more complicated.

\subsection{The geometric groups}
Let us denote by $C_{n}$ the conjugacy classes of $G$. We take the 
GAP labelling. With respect to the geometric subgroups, the classes containing
geometric elements will decompose into conjugacy classes of the $G_i$.

\subsubsection{$G_0=(\z_2^4)\rtimes(\z_2\times\z_2)$}
The group order is $|G_0|=64$, and the generators are $\alpha_1$, $\alpha_2$, $\iota_1$, $\iota_2$, $\iota_3$, $\iota_4$.
Their conjugacy classes sit in those of the octad group as 
\bea
\COne &=& \{ [1]\}\\
\CTwo &\supset& \{ [\iota_2 \iota_1], [\iota_3 \iota_2], [\iota_4 \iota_3 \iota_2 \iota_1], 
[\iota_4 \iota_2 \iota_1], [\iota_1], [\iota_3 \iota_1]\}\\
\CEleven&\supset& \{[ \alpha_1], [\alpha_2], [\alpha_2 \alpha_1]\} \\
\CTwelve&\supset& \{ [\iota_2 \iota_1 \alpha_2] , [  \iota_2 \iota_1 \alpha_2 \alpha_1], [\iota_4 \iota_1 \alpha_1],
[\iota_1 \alpha_2], [\iota_2 \alpha_2 \alpha_1],[\iota_1 \alpha_1] \}\ . 
\eea
From (\ref{z24rep}) we see that all elements in $\CTwo$
have $N_{\rm fp}=-1$. (\ref{repalpha1}) and (\ref{repalpha2})
on the other hand show that
$\alpha_1$, $\alpha_2$, and  $\alpha_2\alpha_1$ all have eigenvalues $\pm i$. 
(\ref{a8rep}) shows that their fixed points are $A,N,O$ and $E,M,N$ and
$D,L,N$ respectively. For $\CEleven$ we thus have $N_{\rm fp}=3$.
For $\CTwelve$ on the other hand we find from the
action of the $\iota_i$ that $N_{\rm fp}=-1$. 
The character for all the conjugacy classes is thus
\be
\phi_g^{\rm tw}(\tau) = \left\{ \begin{array}{cc}   15 \hat\phi_1(\tau) \qquad & g \in \COne\\
 - \hat\phi_1(\tau) \qquad & g\in \CTwo\\
3\hat{\phi}_i (\tau) \qquad & g\in \CEleven\\
-\hat{\phi}_i (\tau) \qquad & g\in \CTwelve
\end{array}\right .
\ee

\subsubsection{$G_1=(\z_2^4)\rtimes A_4$}

The group order is now $|G_1|=192$, and the generators are $\gamma_1$, $\gamma_2$, $\gamma_3$, $\iota_1$, $\iota_2$, $\iota_3$, $\iota_4$. 
Their conjugacy classes now sit inside those of $G$ as 
\bea
\COne&=&\{[1]\}\\
\CTwo&=& \{ [\iota_1], [\iota_2 \iota_1]\}\\
\CSeven&=& \{ [ \gamma_3], [ \gamma_3^{-1}]\}\\
\CEleven&=&\{ [ \gamma_1]\}\\
\CTwelve&=&\{ [\iota_1 \gamma_1] , [\iota_4 \gamma_2], [\iota_1 \gamma_2]\} \ . 
\eea
Noting again that $\gamma_1$ and $\gamma_2$ have
eigenvalues $\pm i$ and that $\gamma_3$ and $\gamma_3^{-1}$
have no fixed points, we get
\be
\phi_g^{\rm tw}(\tau) = \left\{ \begin{array}{cc}   15 \hat\phi_1(\tau) \qquad & g \in \COne\\
 - \hat\phi_1(\tau)\qquad & g\in \CTwo\\
0\qquad & g\in \CSeven\\
3\hat{\phi}_i(\tau) \qquad & g\in \CEleven\\
-\hat{\phi}_i(\tau) \qquad & g\in \CTwelve
\end{array}\right .
\ee

\subsubsection{$G_2=(\z_2^4)\rtimes S_3$}

The group order of $G_2$ is $|G_2|=96$, and the group is generated by  $\beta_1$, $\beta_2$, $\iota_1$, $\iota_2$, $\iota_3$,
and $\iota_4$. The conjugacy classes of $G_2$ sit in those of the octad group as 
\bea
\COne&=&\{[1]\}\\
\CTwo&=& \{ [\iota_1], [\iota_3 \iota_2] , [\iota_3 \iota_1], [\iota_4 \iota_2]\}\\
\CSeven&=& \{ [\beta_1]\}\\
\CEleven&=& \{ [\beta_2]\}\\
\CTwelve&=& \{ [ \iota_2 \iota_1 \beta_2], [\iota_1 \beta_2], [\iota_2 \beta_2]\} \ . 
\eea
Again, $\beta_2$ has eigenvalues $\pm i$,
and $\beta_1$ has no fixed point, so that
\be
\phi_g^{\rm tw}(\tau) = \left\{ \begin{array}{cc}   15 \hat\phi_1(\tau) \qquad & g \in \COne\\
 - \hat\phi_1(\tau) \qquad & g\in \CTwo\\
0 \qquad & g\in \CSeven\\
3\hat{\phi}_i (\tau)  \qquad & g\in \CEleven\\
-\hat{\phi}_i(\tau) \qquad & g\in \CTwelve 
\end{array}\right .
\ee

\subsection{Geometric action on $\Vrd$}\label{app:geocharsrest}
Since the result only depends on the eigenvalues
of $g$, it is straightforward to compute the character
of $\Vrd$ in eq.~(\ref{phir}), using (\ref{phihatu}) and (\ref{phiut}). We find
\be
\phi_g^{r}(\tau) = \left\{ \begin{array}{cc}   -8\frac{\vartheta_4(\tau)^2}{\vartheta_2(\tau)^2} + \tfrac{1}{2} \frac{\vartheta_2(\tau)^2}{\vartheta_4(\tau)^2}
	\qquad & g \in \COne,\CTwo\\[4pt]
	\phi_\zeta^{\rm ut}(\tau) + 
	\hat\phi_\zeta(\tau) \qquad & g\in \CSeven\\[4pt]
	-4 \frac{\vartheta_3(2\tau)}{\vartheta_2(2\tau)} 
	 + \frac{ \vartheta_2(2\tau)}{\vartheta_3(2\tau)} \qquad & g\in \CEleven, \CTwelve 
\end{array}\right .
\ee

\begin{landscape}
\section{The explicit decomposition for the first few levels}\label{app:D}

\bgroup
\def\arraystretch{1.1}
\setlength\tabcolsep{6pt}
\begin{table}[h]%
{\small
\begin{tabular}{|l|c||c|c|}
\hline
$h$&$\vm_h$&$\vo_h$&$\vr_h$\\\hline
1&$\textbf{45}\oplus\con{45}$&$\textbf{45}'\oplus\con{45}'$&$0$\\

2&$\textbf{231}\oplus\con{231}$&$\textbf{210}\oplus\textbf{210}$&$\textbf{21}'\oplus\con{21}'$\\

3&$\textbf{770}\oplus\con{770}$&$2\cdot(\textbf{105}_2\oplus\textbf{315}_1\oplus\textbf{315}_2)$&$\textbf{35}\oplus\textbf{35}$\\

\multirow{2}{*}{$4$}	& \multirow{2}{*}{$\textbf{2277}\oplus\textbf{2277}$}	& $2\cdot(\textbf{15}\oplus\textbf{90}\oplus\textbf{105}_1\oplus\textbf{105}_3\oplus\textbf{120}\oplus 2\cdot\textbf{210}$		& \multirow{2}{*}{$2\cdot (\textbf{28}\oplus\textbf{64}\oplus\textbf{70})$}\\
   && $\oplus~2\cdot\textbf{315}_1\oplus 2\cdot\textbf{315}_2)$&\\
	
\multirow{2}{*}{$5$}	& \multirow{2}{*}{$2\cdot\textbf{5796}$}	& $2\cdot(2\cdot\textbf{45}'\oplus 2\cdot\con{45}'\oplus\textbf{90}\oplus\textbf{105}_1\oplus 2\cdot\textbf{105}_2\oplus\textbf{105}_3\oplus 3\cdot\textbf{120}$		& \multirow{2}{*}{$2\cdot (\textbf{45}\oplus\con{45}\oplus 2\cdot\textbf{56}\oplus\textbf{64}\oplus\textbf{70})$}\\
   && $\oplus~3\cdot\textbf{210}\oplus 6\cdot\textbf{315}_1\oplus 6\cdot\textbf{315}_2)$&\\

\multirow{2}{*}{$6$}	& \multirow{2}{*}{$2\cdot (\textbf{3520}\oplus\textbf{10395})$}	& $2\cdot(\textbf{15}\oplus\textbf{45}'\oplus\con{45}'\oplus 5\cdot\textbf{90}\oplus 6\cdot\textbf{105}_1\oplus 4\cdot\textbf{105}_2\oplus 6\cdot\textbf{105}_3$		& $2\cdot (\textbf{7}\oplus\textbf{14}\oplus\textbf{20}\oplus 2\cdot\textbf{21}\oplus\textbf{21}'\oplus\con{21}'\oplus 2\cdot\textbf{28}\oplus\textbf{35}$\\
   && $\oplus~4\cdot\textbf{120}\oplus 10\cdot\textbf{210}\oplus 13\cdot\textbf{315}_1\oplus 13\cdot\textbf{315}_2)$&$\oplus~2\cdot\textbf{45}\oplus 2\cdot\con{45}\oplus 2\cdot\textbf{56}\oplus 3\cdot\textbf{64}\oplus 3\cdot\textbf{70}$)\\

\hline
\end{tabular}}
\caption{On the left side, we show the decomposition of the coefficients $A_h$ into $\mg$ representations for $h=1,\ldots,6$. The two right columns display how these representations split into $V^\bot_h$ and $V^{\rm rest}_h$ over the Octad subgroup $G$.}
\label{tab:splitting1}
\end{table}
}

\end{landscape}

\begin{landscape}

\section{Character Tables}{\label{app:chartable}}

\begin{table}[h]
	{\scriptsize
		
\begin{tabular}{|c|c|c|c|c|c|c|c|c|c|c|c|c|c|c|c|c|c|c|c|c|c|c|c|c|c|}
\hline
 & 1a & 2a& 3a& 5a& 15a& 15b& 2b& 4a& 4b& 8a& 6a&  2c& 2d& 4c& 4d& 4e& 4f& 3b& 6b& 6c& 12a& 7a& 14a& 7b& 14b\\
\hline
\bf{1}&1&1&1&1&1&1&1&1&1&1&1&1&1&1&1&1&1&1&1&1&1&1&1&1&1\\
\hline
\bf{7}&7&7&4&2&-1&-1&3&3&1&1&.&-1&-1&-1&-1&-1&-1&1&1&-1&-1&.&.&.&.\\
\hline
\bf{14}&14&14&-1&-1&-1&-1&2&2&.&.&-1&6&6&6&2&2&2&2&2&.&.&.&.&.&.\\
\hline
\bf{15}&15&-1&.&.&.&.&3&-1&1&-1&.&7&-1&-1&3&-1&-1&3&-1&1&-1&1&-1&1&-1\\
\hline
\bf{20}&20&20&5&.&.&.&4&4&.&.&1&4&4&4&.&.&.&-1&-1&1&1&-1&-1&-1&-1\\
\hline
\bf{21}&21&21&6&1&1&1&1&1&-1&-1&-2&-3&-3&-3&1&1&1&.&.&.&.&.&.&.&.\\
\hline
$\mathbf{21}'$&21&21&-3&1&$A$&$A^*$&1&1&-1&-1&1&-3&-3&-3&1&1&1&.&.&.&.&.&.&.&.\\
\hline
$\mathbf{\overline{21}'}$&21&21&-3&1&$A^*$&$A$&1&1&-1&-1&1&-3&-3&-3&1&1&1&.&.&.&.&.&.&.&.\\
\hline
\bf{28}&28&28&1&-2&1&1&4&4&.&.&1&-4&-4&-4&.&.&.&1&1&-1&-1&.&.&.&.\\
\hline
\bf{35}&35&35&5&.&.&.&-5&-5&-1&-1&1&3&3&3&-1&-1&-1&2&2&.&.&.&.&.&.\\
\hline
\bf{45}&45&45&.&.&.&.&-3&-3&1&1&.&-3&-3&-3&1&1&1&.&.&.&.&$B$&$B$&$B^*$&$
B^*$\\
\hline
$\mathbf{\overline{45}}$&45&45&.&.&.&.&-3&-3&1&1&.&-3&-3&-3&1&1&1&.&.&.&.&$B^*$&$B^*$&$B$&$B$\\
\hline
$\mathbf{45'}$&45&-3&.&.&.&.&-3&1&1&-1&.&-3&5&-3&1&-3&1&.&.&.&.&$B$&$-B$&$B^*$&$-B^*$\\
\hline
$\mathbf{\overline{45}'}$&45&-3&.&.&.&.&-3&1&1&-1&.&-3&5&-3&1&-3&1&.&.&.&.&$B^*$&$-B^*$&$B$&$-B$\\
\hline
\bf{56}&56&56&-4&1&1&1&.&.&.&.&.&8&8&8&.&.&.&-1&-1&-1&-1&.&.&.&.\\
\hline
\bf{64}&64&64&4&-1&-1&-1&.&.&.&.&.&.&.&.&.&.&.&-2&-2&.&.&1&1&1&1\\
\hline
\bf{70}&70&70&-5&.&.&.&2&2&.&.&-1&-2&-2&-2&-2&-2&-2&1&1&1&1&.&.&.&.\\
\hline
\bf{90}&90&-6&.&.&.&.&6&-2&.&.&.&18&2&-6&2&2&-2&.&.&.&.&-1&1&-1&1\\
\hline
$\mathbf{105}_1$&105&-7&.&.&.&.&-3&1&-1&1&.&17&-7&1&1&-3&1&3&-1&-1&1&.&.&.&.\\
\hline
$\mathbf{105}_2$&105&-7&.&.&.&.&-3&1&-1&1&.&1&9&-7&-3&1&1&3&-1&1&-1&.&.&.&.\\
\hline
$\mathbf{105}_3$&105&-7&.&.&.&.&9&-3&1&-1&.&-7&1&1&-3&1&1&3&-1&-1&1&.&.&.&.\\
\hline
\bf{120}&120&-8&.&.&.&.&.&.&.&.&.&8&8&-8&.&.&.&-3&1&-1&1&1&-1&1&-1\\
\hline
\bf{210}&210&-14&.&.&.&.&6&-2&.&.&.&10&-6&2&-2&-2&2&-3&1&1&-1&.&.&.&.\\
\hline
$\mathbf{315}_1$&315&-21&.&.&.&.&-9&3&1&-1&.&3&-5&3&-1&3&-1&.&.&.&.&.&.&.&.\\
\hline
$\mathbf{315}_2$&315&-21&.&.&.&.&3&-1&-1&1&.&-21&3&3&3&-1&-1&.&.&.&.&.&.&.&.\\
\hline
\end{tabular}
}
\caption{Character table of $G$. $A= \frac{-1-i\sqrt{15}}{2}$, $B= \frac{-1-i\sqrt{7}}{2}$.\label{t:charTableG}. The dots represent $0$.}
\end{table}

\begin{table}[h]
	{\scriptsize
		\setlength\tabcolsep{4.8pt}
		\begin{tabular}{|c|c|c|c|c|c|c|c|c|c|c|c|c|c|c|c|c|c|c|c|c|c|c|c|c|c|c|c|c|c|c|c|c|}
			\hline
			&$1\mathbf{a}$&$2\mathbf{a}$&$2\mathbf{b}$&$2\mathbf{c}$&$2\mathbf{d}$&$4\mathbf{a}$&$2\mathbf{e}$&$4\mathbf{b}$&$2\mathbf{f}$&$4\mathbf{c}$&$2\mathbf{g}$&$4\mathbf{d}$&$4\mathbf{e}$&$4\mathbf{f}$&$4\mathbf{g}$&$4\mathbf{h}$&$2\mathbf{h}$&$6\mathbf{a}$&$3\mathbf{a}$&$6\mathbf{b}$&$12\mathbf{a}$&$6\mathbf{c}$&$4\mathbf{i}$&$4\mathbf{j}$&$4\mathbf{k}$&$8\mathbf{a}$&$4\mathbf{l}$&$4\mathbf{m}$&$7\mathbf{a}$&$14\mathbf{a}$&$7\mathbf{b}$&$14\mathbf{b}$\\ \hline
			
			$\mathbf{1}$& 1& 1& 1& 1& 1& 1& 1& 1& 1& 1& 1& 1& 1& 1& 1& 1& 1& 1& 1& 1& 1& 1& 1& 1& 1& 1& 1& 1& 1& 1& 1& 1 \\ \hline
			
			$\mathbf{1'}$& 1& 1& 1& 1& -1& -1& 1& 1& 1& 1& 1& 1& 1& -1& -1& -1& -1& 1& 1& 1& -1& -1& 1& 1& 1& -1& -1& -1& 1& -1& 1& -1  \\ \hline 
			
			$\mathbf{3}$& 3& 3& 3& 3& -3& -3& -1& -1& -1& -1& -1& -1& -1& 1& 1& 1& 1& .& .& .& .& .& 1& 1& 1& -1& -1& -1& $B^*$& $-B^*$& $B$&$-B$  \\ \hline
			
			$\mathbf{\overline{3}}$&  3& 3& 3& 3& -3& -3& -1& -1& -1& -1& -1& -1& -1& 1& 1& 1& 1& .& .& .& .& .& 1& 1& 1& -1& -1& -1&$B$&$-B$&$B^*$&$-B^*$  \\ \hline
			
			$\mathbf{3'}$&   3& 3& 3& 3& 3& 3& -1& -1& -1& -1& -1& -1& -1& -1& -1& -1& -1& .& .& .& .& .& 1& 1& 1& 1& 1& 1&  $B^*$& $B^*$& $B$& $B$  \\ \hline
			
			$\mathbf{\overline{3}'}$&   3& 3& 3& 3& 3& 3& -1& -1& -1& -1& -1& -1& -1& -1& -1& -1& -1& .& .& .& .& .& 1& 1& 1& 1& 1& 1& $B$& $B$& $B^*$& $B^*$  \\ \hline
			
			$\mathbf{6}$&  6& 6& 6& 6& 6& 6& 2& 2& 2& 2& 2& 2& 2& 2& 2& 2& 2& .& .& .& .& .& .& .& .& .& .& .& -1& -1& -1& -1  \\ \hline 
			
			$\mathbf{6'}$&   6& 6& 6& 6& -6& -6& 2& 2& 2& 2& 2& 2& 2& -2& -2& -2& -2& .& .& .& .& .& .& .& .& .& .& .& -1& 1&-1& 1  \\ \hline 
			
			$\mathbf{7}_1$&   7& 7& 7& 7& 7& 7& -1& -1& -1& -1& -1& -1& -1& -1& -1& -1& -1& 1& 1& 1& 1& 1& -1& -1& -1& -1& -1&-1& .& .& .& .  \\ \hline
			
			$\mathbf{7}_2$&      7& 7& 7& 7& -7& -7& -1& -1& -1& -1& -1& -1& -1& 1& 1& 1& 1& 1& 1& 1& -1& -1& -1& -1& -1& 1& 1& 1& .& .& .& .  \\ \hline
			
			$\mathbf{7}_3$&   7& -1& 7& -1& 7& -1& -1& -1& 3& -1& 3& -1& 3& -1& -1& 3& 3& -1& 1& 1& -1& 1& -1& 1& 1& -1& 1& 1& .& .& .& .  \\ \hline
			
			$\mathbf{7}_4$&    7& -1& 7& -1& -7& 1& -1& -1& 3& -1& 3& -1& 3& 1& 1& -3& -3& -1& 1& 1& 1& -1& -1& 1& 1& 1& -1& -1& .& .& .& .  \\ \hline
			
			$\mathbf{7}_5$&   7& -1& 7& -1& -7& 1& 3& -1& -1& -1& -1& 3& -1& -3& 1& 1& 1& -1& 1& 1& 1& -1& 1& -1& -1& -1& 1& 1& .& .& .& .  \\ \hline
			
			$\mathbf{7}_6$&     7& -1& 7& -1& 7& -1& 3& -1& -1& -1& -1& 3& -1& 3& -1& -1& -1& -1& 1& 1& -1& 1& 1& -1& -1& 1& -1& -1& .& .& .& .  \\ \hline
			
			$\mathbf{8}$&    8& 8& 8& 8& 8& 8& .& .& .& .& .& .& .& .& .& .& .& -1& -1& -1& -1& -1& .& .& .& .& .& .& 1& 1& 1&1  \\ \hline 
			
			$\mathbf{8'}$&    8& 8& 8& 8& -8& -8& .& .& .& .& .& .& .& .& .& .& .& -1& -1& -1& 1& 1& .& .& .& .& .& .& 1& -1& 1& -1  \\ \hline 
			
			$\mathbf{14}_1$&    14& -2& 14& -2& 14& -2& 2& -2& 2& -2& 2& 2& 2& 2& -2& 2& 2& 1& -1& -1& 1& -1& .& .& .& .& .& .&  .& .& .& .  \\ \hline
			
			$\mathbf{14}_2$&   14& -2& 14& -2& -14& 2& 2& -2& 2& -2& 2& 2& 2& -2& 2& -2& -2& 1& -1& -1& -1& 1& .& .& .& .& .& .& .& .& .& .  \\ \hline
			
			$\mathbf{14}_3$&  14& 6& -2& -2& .& .& 2& -2& -2& 2& 6& -2& -2& .& .& .& .& .& 2& -2& .& .& .& -2& 2& .& .& .& .& .& .& .  \\ \hline 
			
			$\mathbf{14}_4$&   14& 6& -2& -2& .& .& 2& 2& 6& -2& -2& -2& -2& .& .& .& .& .& 2& -2& .& .& .& 2& -2& .& .& .& .& .& .& .  \\ \hline 
			
			$\mathbf{21}_1$&   21& -3& 21& -3& 21& -3& -3& 1& 1& 1& 1& -3& 1& -3& 1& 1& 1& .& .& .& .& .& 1& -1& -1& 1& -1& -1& .& .& .& .  \\ \hline
			
			$\mathbf{21}_2$&    21& -3& 21& -3& -21& 3& -3& 1& 1& 1& 1& -3& 1& 3& -1& -1& -1& .& .& .& .& .& 1& -1& -1& -1& 1& 1&.& .& .& .  \\ \hline
			
			$\mathbf{21}_3$&    21& -3& 21& -3& -21& 3& 1& 1& -3& 1& -3& 1& -3& -1& -1& 3& 3& .& .& .& .& .& -1& 1& 1& 1& -1& -1&  .& .& .& .  \\ \hline
			
			$\mathbf{21}_4$&   21& -3& 21& -3& 21& -3& 1& 1& -3& 1& -3& 1& -3& 1& 1& -3& -3& .& .& .& .& .& -1& 1& 1& -1& 1& 1&  .& .& .& .  \\ \hline
			
			$\mathbf{28}$&   28& 12& -4& -4& .& .& 4& .& 4& .& 4& -4& -4& .& .& .& .& .& -2& 2& .& .& .& .& .& .& .& .& .& .& .& .  \\ \hline 
			
			$\mathbf{42}_1$&   42& 18& -6& -6& .& .& -2& -2& -6& 2& 2& 2& 2& .& .& .& .& .& .& .& .& .& .& 2& -2& .& .& .& .& .&  .& .  \\ \hline 
			
			$\mathbf{42}_2$&   42& -6& -6& 2& .& .& -2& 2& -2& -2& 6& 2& -2& .& .& -4& 4& .& .& .& .& .& .& .& .& .& -2& 2& .& .& .& .  \\ \hline 
			
			$\mathbf{42}_3$&  42& -6& -6& 2& .& .& -2& 2& -2& -2& 6& 2& -2& .& .& 4& -4& .& .& .& .& .& .& .& .& .& 2& -2& .&  .& .& .  \\ \hline 
			
			$\mathbf{42}_4$&   42& 18& -6& -6& .& .& -2& 2& 2& -2& -6& 2& 2& .& .& .& .& .& .& .& .& .& .& -2& 2& .& .& .& .& .&  .& .  \\ \hline 
			
			$\mathbf{42}_5$&    42& -6& -6& 2& .& .& -2& -2& 6& 2& -2& 2& -2& .& .& -4& 4& .& .& .& .& .& .& .& .& .& 2& -2& .&  .& .& .  \\ \hline 
			
			$\mathbf{42}_6$&    42& -6& -6& 2& .& .& -2& -2& 6& 2& -2& 2& -2& .& .& 4& -4& .& .& .& .& .& .& .& .& .& -2& 2& .&  .& .& .  \\ \hline 
			
			$\mathbf{84}$&    84& -12& -12& 4& .& .& 4& .& -4& .& -4& -4& 4& .& .& .& .& .& .& .& .& .& .& .& .& .& .& .& .& .& .& .  \\ \hline \hline
			
			&1a&2c&2a&2d&2a&4c&2b&4d&2c&4e&2d&4a&4c&4a&4f&4c&2d&6c&3b&6b&12a&6b&4b&4d&4f&8a&4e&4f&7b&14b&7a&14a\\ \hline
	\end{tabular}}
	\caption{Character table of $\HA=((\z_2)^6\rtimes PSL(3,2))\rtimes\z_2$. $\HA$ is a subgroup of $G$ of order $21504$. The bottom row indicates which conjugacy class of $G$ the classes of $\HA$ belong to.}
	\label{t:charTableH}
\end{table}

\end{landscape}

\bibliographystyle{ieeetr}
\bibliography{ref}

\end{document}